\documentclass[epj]{svjour}
\usepackage[utf8]{inputenc}
\usepackage{graphicx}
\usepackage{color}
\usepackage{amsfonts}
\usepackage{amssymb}
\usepackage{amsmath}
\usepackage{widetext}
\newcommand{\be}{\begin{equation}} 
\newcommand{\ee}{\end{equation}}
\newcommand{\bea}{\begin{eqnarray}} 
\newcommand{\eea}{\end{eqnarray}}
\newcommand{\nn}{\nonumber}

\begin{document}

\title{Phenomenology of Self-Interacting Dark Matter in a Matter-Dominated Universe}
\titlerunning{Phenomenology of Self-Interacting Dark Matter}
\author{Nicolás Bernal\inst{1}, Catarina Cosme\inst{2} \and Tommi Tenkanen\inst{3,4}}

\institute{Centro de Investigaciones, Universidad Antonio Nari\~{n}o, Carrera 3 Este \# 47A-15, Bogot\'{a}, Colombia \and
Departamento de F\'{\i}sica e Astronomia, Faculdade de Ci\^encias da Universidade do Porto and Centro de F\'{\i}sica do Porto,\\Rua do Campo Alegre 687, 4169-007, Porto, Portugal\and
Department of Physics and Astronomy, Johns Hopkins University, Baltimore, MD 21218, United States of America \and
Astronomy Unit, Queen Mary University of London, Mile End Road, London, E1 4NS, United Kingdom}

\date{Received: date / Revised version: date}

\abstract{
We study production of self-interacting dark matter (DM) during an early matter-dominated phase. As a benchmark scenario, we consider a model where the DM consists of singlet scalar particles coupled to the visible Standard Model (SM) sector via the Higgs portal. We consider scenarios where the initial DM abundance is set by either the usual thermal {\it freeze-out} or an alternative {\it freeze-in} mechanism, where DM was never in thermal equilibrium with the SM sector. For the first time, we take the effect of self-interactions within the hidden sector into account in determining the DM abundance, reminiscent to the Strongly Interacting Massive Particle (SIMP) scenario. In all cases, the number density of DM may change considerably compared to the standard radiation-dominated case, having important observational and experimental ramifications.
}

\maketitle



\section{Introduction}
\label{intro}

The existence of dark matter (DM) seems indisputable. From the Cosmic Microwave Background radiation (CMB), large scale structure of the Universe and different physics at galactic scales, one can infer that there must be a long-lived, dynamically non-hot, non-baryonic matter component, whose abundance exceeds the amount of ordinary `baryonic' matter roughly by a factor of five~\cite{Bergstrom:2000pn,Aghanim:2018eyx,Bertone:2016nfn,deSwart:2017heh} and which has been there from the hot Big Bang era until the present day. However, the non-gravitational nature of the DM component remains a mystery.

For a long time, Weakly Interacting Massive Particles (WIMPs) have been among the best-motivated DM candidates. The increasingly strong observational constraints on DM (see e.g. Ref.~\cite{Arcadi:2017kky}) are, however, not only puzzling as such but are now forcing one to ask: is the standard WIMP paradigm just waning, or is it already dead? If so, what alternative explanations for the production and properties of DM do we have?

A simple alternative for the standard WIMPs is provided by relaxing the usual assumption that DM is a thermal relic, produced by the {\it freeze-out} (FO) mechanism in the early Universe, and assuming that it never entered in thermal equilibrium with the particles within the Standard Model of particle physics (SM). If that was the case, then the present DM abundance could have been produced by the so-called {\it freeze-in} (FI) mechanism, where the abundance results from decays and annihilations of SM particles into DM~\cite{McDonald:2001vt,Choi:2005vq,Kusenko:2006rh,Petraki:2007gq,Hall:2009bx}. Assuming that DM never entered into thermal equilibrium with the particles in the visible SM sector typically amounts to choosing a very small coupling between the two sectors. A good thing about this is that then these so-called Feebly Interacting Massive Particles (FIMPs) easily evade the increasingly stringent observational constraints, yet an obvious hindrance is that this also makes the scenario inherently very difficult to test. For a recent review of FIMP DM models and observational constraints presented in the literature, see Ref.~\cite{Bernal:2017kxu}.

Another way to evade the experimental constraints is to consider non-standard cosmological histories~\cite{Kamionkowski:1990ni}. We know that the Universe was effectively radiation-dominated (RD) at the time of Big Bang Nucleosynthesis (BBN) and one usually assumes that this was the case also at the time the DM component was produced, was it at the time of electroweak cross-over or at higher energy scales. However, there are no obvious reasons for limiting the DM studies on such cosmological expansion histories,\footnote{A possible caveat to this is the viability of models for baryogenesis in such scenarios. However, some studies have shown that baryogenesis with a low reheating temperature may be much less difficult than expected~\cite{Davidson:2000dw,Giudice:2000ex,Allahverdi:2010im,Allahverdi:2017edd}. Furthermore, there are some baryogenesis scenarios with MD cosmologies~\cite{Bernal:2017zvx}.} as alternatives not only can lead to interesting observational ramifications but are also well-motivated. For example, an early matter-dominated (MD) phase can be caused by late-time reheating \cite{Allahverdi:2010xz}, massive meta-stable particles governing the energy density of the Universe (see Refs.~\cite{Berlin:2016vnh,Tenkanen:2016jic,Berlin:2016gtr} for recent works), moduli fields~\cite{Vilenkin:1982wt,Starobinsky:1994bd,Dine:1995uk}, and so on. The effect on the resulting DM yield can then be outstanding, as recently studied in detail in e.g. Refs.~\cite{Co:2015pka,Berlin:2016vnh,Tenkanen:2016jic,Dror:2016rxc,Berlin:2016gtr,DEramo:2017gpl,Hamdan:2017psw,Drees:2017iod,Dror:2017gjq,DEramo:2017ecx,Hardy:2018bph}.

Indeed, when the expansion rate of the Universe differs from the usual RD case, it tends to effectively dilute the DM abundance when the era of non-standard expansion ends and the visible sector gets reheated (see also Refs.~\cite{DEramo:2017gpl,DEramo:2017ecx} for DM production in fast-expanding universes and Refs.~\cite{Dror:2016rxc,Dror:2017gjq} for co-decaying DM). This means, for example, that when the expansion was faster than in the RD case and the DM particles were initially in thermal equilibrium with the visible sector, they generically have to undergo freeze-out {\it earlier} than in the usual RD case, thus resulting in larger DM abundance to match the observed one. In case the DM particles interacted so feebly that they were never part of the equilibrium heat bath, the coupling between DM and the visible sector typically has to be orders of magnitude larger than in the usual freeze-in case to compensate the larger expansion rate. Production of DM during a non-standard expansion phase may thus result to important experimental and observational ramifications. Studying the effect non-standard cosmological histories have on different particle physics scenarios is thus not only of academic interest and also not limited to the final DM abundance, as different possibilities to test for example an early MD phase include formation of ultracompact substructures such as microhalos \cite{Erickcek:2011us} or primordial black holes \cite{Carr:2017edp,Cole:2017gle,Kohri:2018qtx}, as well as cosmological phase transitions with observational gravitational wave signatures \cite{Barenboim:2016mjm} (see also Ref.~\cite{Beniwal:2017eik}).

In this paper we will consider DM production during such an early MD phase.
We will study DM production by both the freeze-out and freeze-in mechanisms, taking for the first time into account the effect that non-vanishing DM self-interactions can have. Instead of performing an intensive full-parameter scan, in this paper we will perform an analytical study of the different representative cases previously mentioned, which allows us to capture the essence of each scenario. Results of an exhaustive scan over the full parameter space in the usual freeze-out and freeze-in cases are presented in a companion paper~\cite{Bernal:2018kcw}, where we also discuss the effect of other non-standard cosmological histories. However, as we will show, already with the best-motivated non-standard case, an early phase of matter-domination, the DM phenomenology is very rich when the effect of DM self-interactions is taken into account, which is one of the reasons why we devote a separate paper for the analysis of this scenario only. Another important difference to Ref.~\cite{Bernal:2018kcw} is that in this paper we will we make the usual assumption that the eventual decay of the energy density component responsible for the early matter-domination is instantaneous, whereas in Ref.~\cite{Bernal:2018kcw} the duration of decay is taken to be finite. In this way, the two studies complement each other.

As we will show, the observational limits on DM self-interactions do not only rule out part of the parameter space for the model we will consider in this paper, but taking the detailed effect of DM self-interactions into account is crucial for determination of the final DM abundance, reminiscent to the so-called Strongly Interacting Massive Particle (SIMP) or {\it cannibal} DM scenarios~\cite{Dolgov:1980uu,Carlson:1992fn,Hochberg:2014dra,Bernal:2015bla,Bernal:2015lbl,Bernal:2015ova,Bernal:2015xba,Pappadopulo:2016pkp,Heikinheimo:2016yds,Farina:2016llk,Chu:2016pew,Dey:2016qgf,Bernal:2017mqb,Choi:2017mkk,Heikinheimo:2017ofk,Ho:2017fte,Dolgov:2017ujf,Garcia-Cely:2017qpx,Hansen:2017rxr,Chu:2017msm,Duch:2017khv,Chauhan:2017eck,Herms:2018ajr,Heikinheimo:2018esa}. We will also discuss other prospects for detection of DM including collider, direct, and indirect detection experiments.

The paper is organized as follows: In Section~\ref{model}, we will present a simple benchmark model where the DM particle is a real singlet scalar odd under a discrete $\mathbb{Z}_2$ symmetry, and discuss what are the requirements for having an early MD phase prior to BBN. In Section~\ref{DMproduction}, we turn into the DM production, discussing production by the usual freeze-out mechanism in Subsection \ref{sec:freezeout} and by the freeze-in mechanism in Subsection~\ref{sec:freezein}. In Section~\ref{constraints}, we discuss the experimental and observational ramifications, and present not only what part of the parameter space is already ruled out but also what part of it can be probed in the near future. Finally, we conclude with an outlook in Section~\ref{conclusions}.


\section{The Model}
\label{model}

We study an extension of the SM where on top of the SM matter field content we assume a simple hidden sector consisting of a real singlet scalar $s$. The only interaction between this hidden singlet sector and the visible SM sector is via the Higgs portal coupling $\lambda_{hs}|\Phi|^2s^2$, where $\Phi$ is the SM Higgs field. The scalar potential is
\be
\label{potential}
V(\Phi,s) = \mu_h^2|\Phi|^2+\lambda_h|\Phi|^4+\frac{\mu_s^2}{2} s^2+\frac{\lambda_s}{4}s^4+\frac{\lambda_{hs}}{2}|\Phi|^2 s^2 ,
\ee
where $\sqrt{2}\Phi^{\rm T}=(0,v+h)$ is the SM $SU(2)$ gauge doublet in the unitary gauge and $v=246$~GeV is the vacuum expectation value of the SM Higgs field. A discrete $\mathbb{Z}_2$ symmetry, under which the DM is odd and the whole SM is even, has been assumed to stabilize the singlet scalar and make it a possible DM candidate. We assume $\lambda_s>0$ and $\mu_s>0$, so that the minimum of the potential in the $s$ direction is at $s=0$ and $m_s^2 \equiv \mu_s^2 + \lambda_{hs}\,v^2/2$ is the physical mass of $s$ after the spontaneous symmetry breaking in the SM sector. This implies $\lambda_{hs}<2\,m_s^2/v^2$.

\subsection{An Early Matter-dominated Period}
\label{sec:matdom}

We assume that the Universe was MD for the whole duration of DM production down to $T \gtrsim 4$ MeV, where the lower limit is given by BBN~\cite{Kawasaki:2000en,Hannestad:2004px,Ichikawa:2005vw,DeBernardis:2008zz}. By this time, the matter-dominance must have ended, the SM sector must have become the dominant energy density component and the usual Hot Big Bang era must have begun. We assume that when DM was produced, both the SM and the singlet sector were energetically subdominant, so that
\be
\label{friedmann}
3\,H^2 M_{\rm P}^2 = \rho_{\rm total} \simeq \rho_{\rm M} \gg \rho_{\rm SM},\,\rho_s \,,
\ee
where $H$ is the Hubble scale, $M_{\rm P}$ is the reduced Planck mass, and $\rho_{\rm M}$ is the energy density of the matter-like component that is assumed to dominate over the SM energy density $\rho_{\rm SM}$ and the singlet scalar energy density $\rho_s$. We also assume that the SM was in thermal equilibrium for the whole duration of the early MD phase, so that 
\be
\label{SM rho}
\rho_{\rm SM} = \frac{\pi^2}{30}\,g_* \,T^4,
\ee
where $g_*$ is the usual effective number of relativistic degrees of freedom\footnote{In the following sections we will neglect, for simplicity, the evolution of $g_*$ during the DM production. A detailed effect of this is addressed in Ref.~\cite{Bernal:2018kcw}, although the correction this imposes is relatively small.} and $T$ is the SM bath temperature. 

The magnitude of the Hubble expansion rate 
can be understood by first discussing the dynamics in the usual RD case where the SM is the dominant energy density component. In that case, the Friedmann equation~\eqref{friedmann} gives at $T=m_h$ the result
\be
\label{HEW}
\frac{H_{\rm EW}^{\rm rad}}{m_h} = \sqrt{\frac{\pi^2g_*(m_h)}{90}}\frac{m_h}{M_{\rm P}} \simeq 1.76\times 10^{-16} ,
\ee
where we used $g_*(m_h)=106.75$ and denoted $H_{\rm EW}\equiv H(T=m_h)$. However, in a MD Universe at $T=m_h$ we have
\be
3\,H_{\rm EW}^2\,M_{\rm P}^2 = \left.\left(\rho_{\rm M} + \rho_{\rm SM}\right)\right|_{T=m_h} \simeq \left.\rho_{\rm M}\right|_{T=m_h} \,,
\ee
so that in this case $H_{\rm EW}/m_h\gg H_{\rm EW}^{\rm rad}/m_h$, i.e. the Universe expands much faster than in the standard RD case. Determining the ratio $H_{\rm EW}/m_h$ more accurately than this is not possible without specifying the underlying dynamics causing the early MD, so in the remaining of this paper we simply take it to be a free parameter for generality. 

\subsection{Constraints on the Scenario}
\label{sec:constraints}

In all cases, both the model parameters in Eq.~\eqref{potential} and the cosmological parameters are subject to constraints that come from observational data. In this paper, we make the usual assumption that the matter component governing the total energy density decays instantaneously into the SM radiation. The first condition then is that the SM temperature \textit{after} the matter-like component has decayed into SM particles, $T_{\rm end}'$, must be larger than the BBN temperature $T_{\rm BBN}=4$~MeV. Second, the temperature has to be smaller than either the final freeze-out temperature or smaller than $m_h$ in the freeze-in case in order not to re-trigger the DM yield after the decay of the matter-like component. As shown in the end of Appendix~\ref{appendix},
this amounts to requiring
\begin{widetext}
\bea
\label{tend over mh constr}
5\times 10^{-7}\left(\frac{H_{\rm EW}/m_h}{10^{-16}}\right)^{-2/3} \lesssim \frac{T_{\mathrm{end}}}{m_h} &\lesssim
\begin{cases}	 
2\times 10^{-3}\left(\frac{H_{\rm EW}/m_h}{10^{-16}}\right)^{-2/3}\left(\frac{m_s}{\rm GeV}\right)^{4/3}x_{\rm FO}^{-4/3} \quad \text{freeze-out}, \\
\left(\frac{H_{\rm EW}/m_h}{10^{-16}}\right)^{-2/3} \hspace{4.05cm} \text{freeze-in},
\end{cases}
\eea
\end{widetext}
\noindent where $T_{\rm end}$ is the SM temperature just {\it before} the end of matter-domination and $x_{\rm FO}\equiv m_s/T_\text{FO}$, with $T_\text{FO}$ being the DM freeze-out temperature. In the following, we will take the above ratio $T_{\rm end}/m_h$ to be a free parameter, so that together with $H_{\rm EW}$ it constitutes the set of our cosmological parameters, characterizing the duration of the early MD phase. The total parameter space is thus five-dimensional, consisting of the particle physics parameters $\lambda_s$, $\lambda_{hs}$ and $m_s$, in addition to the cosmological parameters $H_\text{EW}/m_h$ and $T_\text{end}/m_h$.

Third, we require that DM freeze-out always occurs while the $s$ particles are non-relativistic, $x_{\rm FO}>3$, as otherwise the scenario is subject to relativistic corrections that we are not taking into account in the present paper. 
Fourth, as discussed above, in a MD Universe $H_{\mathrm{EW}}/m_h\gg 10^{-16}$.
Fifth, as discussed below Eq.~\eqref{potential}, the portal coupling has to satisfy $\lambda_{hs}<2\,m_s^2/v^2$. Finally, the portal coupling has a further constraint when requiring or avoiding the thermalization of the two sectors, for the case of freeze-out and freeze-in, respectively. Depending on the strength of the portal coupling $\lambda_{hs}$, the singlet scalar particles may or may not have been part of the equilibrium in the SM sector at the time the initial DM density was produced. The threshold value for $\lambda_{hs}$ above which the DM sector equilibrates with the SM is
\be
\label{portalcoupling}
\lambda_{hs}^\text{eq} \simeq \sqrt{\frac{128\pi^3}{\zeta(3)}\frac{H_{\rm EW}}{m_h}} \,.
\ee
This results from requiring that the SM particles do not populate the hidden sector so that they would start to annihilate back to the SM in large amounts, $\langle \sigma_{hh \rightarrow ss}v\rangle n_h/H \simeq \lambda_{hs}^2\,\zeta(3)\, m_h/(128\pi^3H_\text{EW}) < 1$ \cite{Enqvist:2014zqa,Alanne:2014bra,Tenkanen:2016twd}, where $\langle\sigma_{hh\to ss}v\rangle$ is the thermally averaged cross-section for the process $hh\to ss$ and $\zeta(3)\simeq 1.20$ is the Riemann zeta function. 
For the freeze-out case we demand $\lambda_{hs}\gg\lambda_{hs}^\text{eq}$ whereas for the freeze-in $\lambda_{hs}\ll\lambda_{hs}^\text{eq}$.

Before concluding this section let us note that the fact that now $H_{\rm EW}\gg H_{\rm EW}^{\rm rad}$ means that in the freeze-out case the value of the portal coupling required to produce the observed DM abundance must be smaller than in the usual RD case, as the DM has to decouple earlier from the thermal bath in order to retain the required abundance. However, the faster expansion rate also means that now the threshold value for thermalization, Eq.~\eqref{portalcoupling}, can be orders of magnitude larger than the corresponding value $\lambda_{hs}\simeq 10^{-7}$ in the usual RD case. This makes the freeze-in scenario particularly interesting, as it might lead to important experimental ramifications, as we will discuss in Section~\ref{constraints}.


\section{Dark Matter Production}
\label{DMproduction}

We start by reviewing the DM production within this model, briefly discussing two
fundamental mechanisms that account for it: the freeze-out and the
freeze-in scenarios.

Assuming that there is only one DM
particle, $s$, its number density evolution is described by the
Boltzmann equation:
\begin{align}
 &\frac{dn_s}{dt}+3\,H\,n_s=\nonumber\\
 & -\int d\Pi_{s}d\Pi_{{\rm a_{1}}}d\Pi_{{\rm a_{2}}}...d\Pi_{b_{1}}d\Pi_{{\rm b_{2}}}...\nonumber \\
 & \times\left(2\pi\right)^{4}\delta^{4}\left(p_{s}+p_{{\rm a_{1}}}+p_{{\rm a_{2}}}...-p_{{\rm b_{1}}}-p_{{\rm b2}}...\right)\nonumber \\
 & \times\left[\left|\mathcal{M}\right|_{\mathrm{s+a_{1}+a_{2}....\rightarrow b_{1}+b_{2}...}}^{2}\,f_{s}\,f_{\mathrm{a_{1}}}...\left(1\pm f_{\mathrm{b_{1}}}\right)\left(1\pm f_{\mathrm{b_{2}}}\right)...\right.\nonumber \\
 & \left.-\left|\mathcal{M}\right|_{\mathrm{b_{1}+b_{2}....\rightarrow s+a_{1}+a_{2}...}\,}^{2}f_{\mathrm{b_{1}}}\,f_{\mathrm{b_{2}}}...\left(1\pm f_{s}\right)\left(1\pm f_{\mathrm{a_{1}}}\right)...\right],\label{Boltz eq general}
\end{align}
considering the process $s+a_{1}+a_{2}+...+a_{k}\rightarrow b_{1}+b_{2}+...+b_{j}$, where $a_i, b_j$ are particles in the heat bath. Here $n_s$ is the DM number density, $p_i$ is the momentum
of the particle $i$, $\left|\mathcal{M}\right|^{2}$ is the squared
transition amplitude averaged over both initial and final states, $f_i$
is the phase space density, $+$ applies to bosons and $-$ to fermions
and 
\begin{equation}
d\Pi_i\equiv\frac{g_i}{\left(2\pi\right)^{3}}\,\frac{d^{3}p_i}{2E_i}\label{phase space}
\end{equation}
is the phase space measure, where $g_i$ is the number of intrinsic degrees of freedom and $E_i$ the energy of the particle $i$. In the following, we will solve the relevant Boltzmann equations analytically in the regions of interest where different processes dominate at a time. A full parameter scan is performed in the pure freeze-out and freeze-in cases in Ref.~\cite{Bernal:2018kcw}.

In the freeze-out mechanism, DM was initially in thermal equilibrium with the SM sector. As soon as the interactions between the DM and the SM particles were no longer able to keep up with the Hubble expansion, the system departed from thermal equilibrium and the comoving DM abundance became constant. We will study the case of the DM freeze-out in an early MD era in Section~\ref{sec:FO w/o cann} and then consider how a so-called cannibalism phase affects the DM yield in Section~\ref{sec:FO w cann}.

In the freeze-in scenario, the DM was never in thermal
equilibrium with the visible sector, due to the very feeble interactions
between them. The particles produced by this mechanism are known as
FIMPs and their initial number density is, in the simplest case, negligible. The DM abundance is
produced by the SM particle decays and annihilations, lasting until the number density
of the SM particles becomes Boltzmann-suppressed. At this point, the comoving number density of DM
particles becomes constant and the comoving DM abundance is said to `freeze in'. The
evolution of the initial $s$ number density can be tracked by the
Boltzmann equation~\eqref{Boltz eq general} as well. We discuss the DM freeze-in in
an early MD era without cannibalism in Section~\ref{sec:FI w/o cann}
and with it in Section~\ref{sec:cannibalism}. 

\subsection{The Freeze-out Case}
\label{sec:freezeout}

To study the effects of MD and DM self-interactions in a simple yet accurate way, in this section we assume the mass hierarchy $m_b<m_s < 50$ GeV, where $m_b$ is the mass of the $b$-quark and the upper limit is chosen to avoid complications with the Higgs resonance in our analytical calculations. Therefore, in this subsection, we will consider DM produced only by $b\bar b$ annihilations and present the more general analysis in Ref.~\cite{Bernal:2018kcw} for the pure freeze-out case without cannibalism.

\subsubsection{Freeze-out without Cannibalism}
\label{sec:FO w/o cann}

In this scenario, we assume that the DM was initially in thermal equilibrium with the SM particles. In the most simple case that we are considering here, only the annihilation
and inverse annihilation processes $ss\leftrightarrow b\bar{b}$
are taken into account for the abundance, and the equation governing the evolution of the
DM number density, \eqref{Boltz eq general}, becomes
\begin{equation}
\frac{dn_{s}}{dt}+3\,H\,n_{s}=-\left\langle \sigma_{ss\rightarrow b\bar{b}} v\right\rangle \left[n_{s}^{2}-\left(n_{s}^\text{eq}\right)^{2}\right],\label{boltzmann FO}
\end{equation}
where $\langle\sigma_{ss\rightarrow b\bar{b}} v\rangle$
is the thermally-averaged DM annihilation cross-section times velocity
and $n_{s}^\text{eq}$ corresponds to the DM equilibrium number
density.

When the interactions between the DM and the visible
sector cannot keep up against the expansion of the Universe any more,
the DM decouples and its comoving number density freezes to a
constant value. This occurs at $T=T_\text{FO}$ defined by
\begin{equation}
\left.\frac{\left\langle \sigma_{ss\rightarrow b\bar{b}}v\right\rangle \,n_{s}}{H}\right|_{T=T_\text{FO}}=1\,.\label{freeze out}
\end{equation}
Assuming that DM is non-relativistic when interactions freeze-out, we have
\begin{equation}
n_{s}(T)=\left(\frac{m_{s}\,T}{2\pi}\right)^{\frac{3}{2}}\,e^{-\frac{m_{s}}{T}},\label{ns equil}
\end{equation}
whereas the Hubble parameter is given by
\begin{equation}
H(T)=H_{\mathrm{EW}}\,\left(\frac{T}{m_h}\right)^{\frac{3}{2}}\,\left(\frac{g_{*}\left(T\right)}{g_{*}\left(m_h\right)}\right)^{\frac{1}{2}}.\label{Hubble}
\end{equation}
Substituting then Eqs. (\ref{ns equil}) and (\ref{Hubble}) into (\ref{freeze out}), the freeze-out condition can be written as
\begin{equation}
x_{\mathrm{FO}}=\ln\left[\frac{\lambda_{hs}^{2}}{2^{9/2}\,\pi^{5/2}}\,\left(\frac{g_{*}\left(m_h\right)}{g_{*}\left(T_{\mathrm{FO}}\right)}\right)^{1/2}\,\left(\frac{H_{\mathrm{EW}}}{m_h}\right)^{-1}\,\frac{m_b^{2}\,m_{s}^{3/2}}{m_h^{7/2}}\right],\label{FO condition}
\end{equation}
where we used $\langle\sigma_{ss\to b\bar{b}}v \rangle \simeq \lambda_{hs}^2m_b^2/(8\pi\, m_h^4)$~\cite{Bernal:2015bla,Bernal:2015lbl} and $x_{\mathrm{FO}}\equiv m_s/T_{\rm FO}$ corresponds
to the time when DM annihilation into $b$-quarks becomes smaller
than the Hubble parameter.
The DM abundance can then be calculated by taking into account the non-conservation of entropy (see Appendix \ref{appendix}), yielding:
\begin{align}
\label{FOresult}
	\frac{\Omega_{s}\,h^{2}}{0.12}\simeq & \,3\times10^{-7}\,x_{{\rm FO}}^{3/2}\,e^{-x_{{\rm FO}}}\,\nonumber\\
	&\times\left(\frac{T_{{\rm end}}}{m_h}\right)^{3/4}\,\left(\frac{H_{\rm EW}/m_h}{10^{-16}}\right)^{-3/2}\,\left(\frac{m_{s}}{{\rm GeV}}\right)\,,
\end{align}
where $x_{{\rm FO}}$ is given by Eq.~\eqref{FO condition}. 
Let us note that in this case, production without cannibalism, the parameter $\lambda_s$ is small ($\lambda_s\lesssim 10^{-3}$) and plays no role in the WIMP DM phenomenology. In the next Subsection we will, however, consider the opposite case where large self-interactions do change the resulting DM abundance.

\begin{figure}
\begin{center}
\includegraphics[width=.49\textwidth]{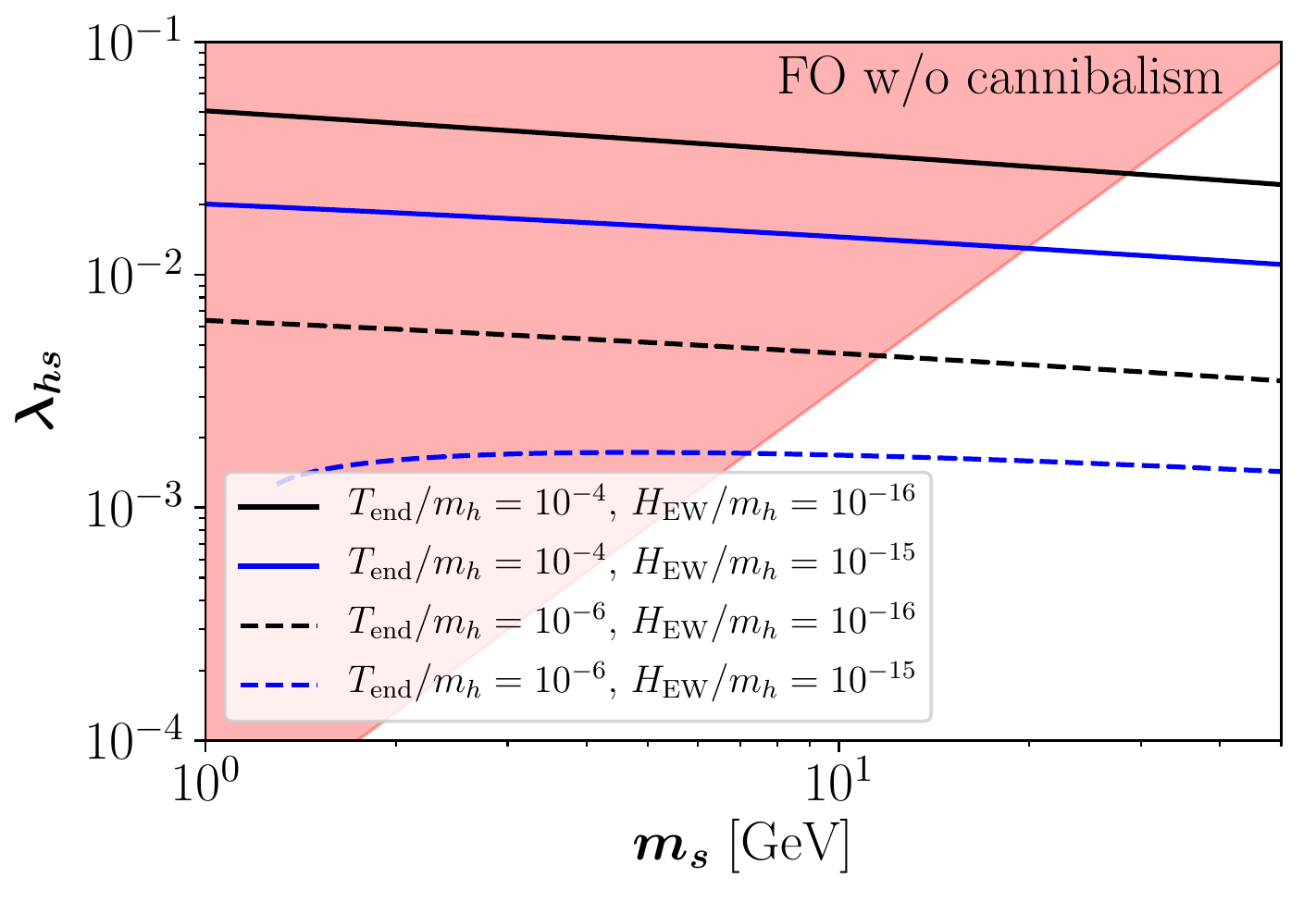}
\includegraphics[width=.49\textwidth]{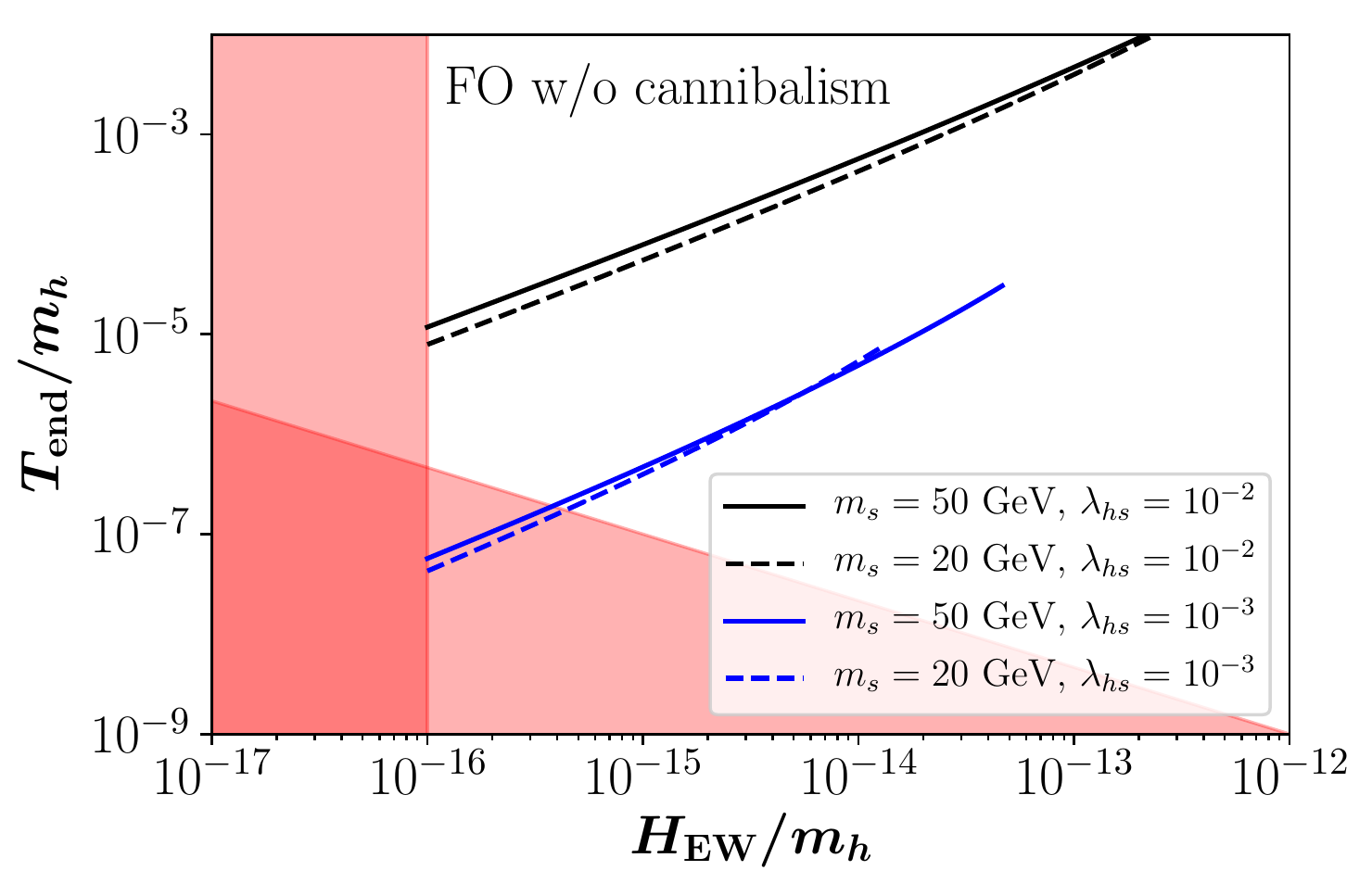}
\caption{DM freeze-out without cannibalism. Parameter space giving rise to the observed DM relic abundance. The red regions correspond to the constraints discussed in Section~\ref{sec:constraints}. Other observational constraints are discussed in Section~\ref{constraints} and shown in Fig.~\ref{Detection}.}
\label{FOwithoutCannibalSlices}
\end{center}
\end{figure}

Fig.~\ref{FOwithoutCannibalSlices} shows slices of the parameter space that give rise to the observed DM relic abundance.
On the upper panel the cosmological parameters are fixed, $H_\text{EW}/m_h=10^{-16}$ (black lines) and $10^{-15}$ (blue lines), and $T_\text{end}/m_h=10^{-6}$ (dashed lines) and $10^{-4}$ (solid lines) while we scan over the relevant particle physics parameters ($\lambda_{hs}$ and $m_s$).
The upper left corner in red, corresponding to $\lambda_{hs}>2\,m_s^2/v^2$, is excluded by the requirement discussed below Eq. \eqref{potential}.
The figure shows that an increase in the dilution factor due to either an enhancement of the Hubble expansion rate $H_\text{EW}$ or a decrease in the temperature $T_\text{end}$ when the MD era ends has to be compensated with a higher DM abundance at the freeze-out.
That, in turn, requires a smaller annihilation cross-section and hence a small $\lambda_{hs}$.
The dependence on the DM mass $m_s$ is very mild.

The same conclusion can be extracted from the lower panel of Fig.~\ref{FOwithoutCannibalSlices}, where the particle physics parameters are fixed, $m_s=20$~GeV (dashed lines) and $50$~GeV (solid lines), and $\lambda_{hs}=10^{-3}$ (blue lines) and $10^{-2}$ (black lines) while we scan over the cosmological parameters.
The left red band corresponds to a scenario which is not MD\linebreak ($H_\text{EW}/m_h<10^{-16}$), whereas the lower left corner corresponds to a case where the resulting SM temperature after the MD era ends is too small for successful BBN. Both cases are excluded from our analysis. Here the requirement of a non-relativistic freeze-out ($x_\text{FO}>3$) is also taken into account. Other observational constraints on the scenario will be discussed in Section \ref{constraints}.

\subsubsection{Freeze-out with Cannibalism}
\label{sec:FO w cann}

\begin{figure}
\begin{center}
\includegraphics[width=.50\textwidth]{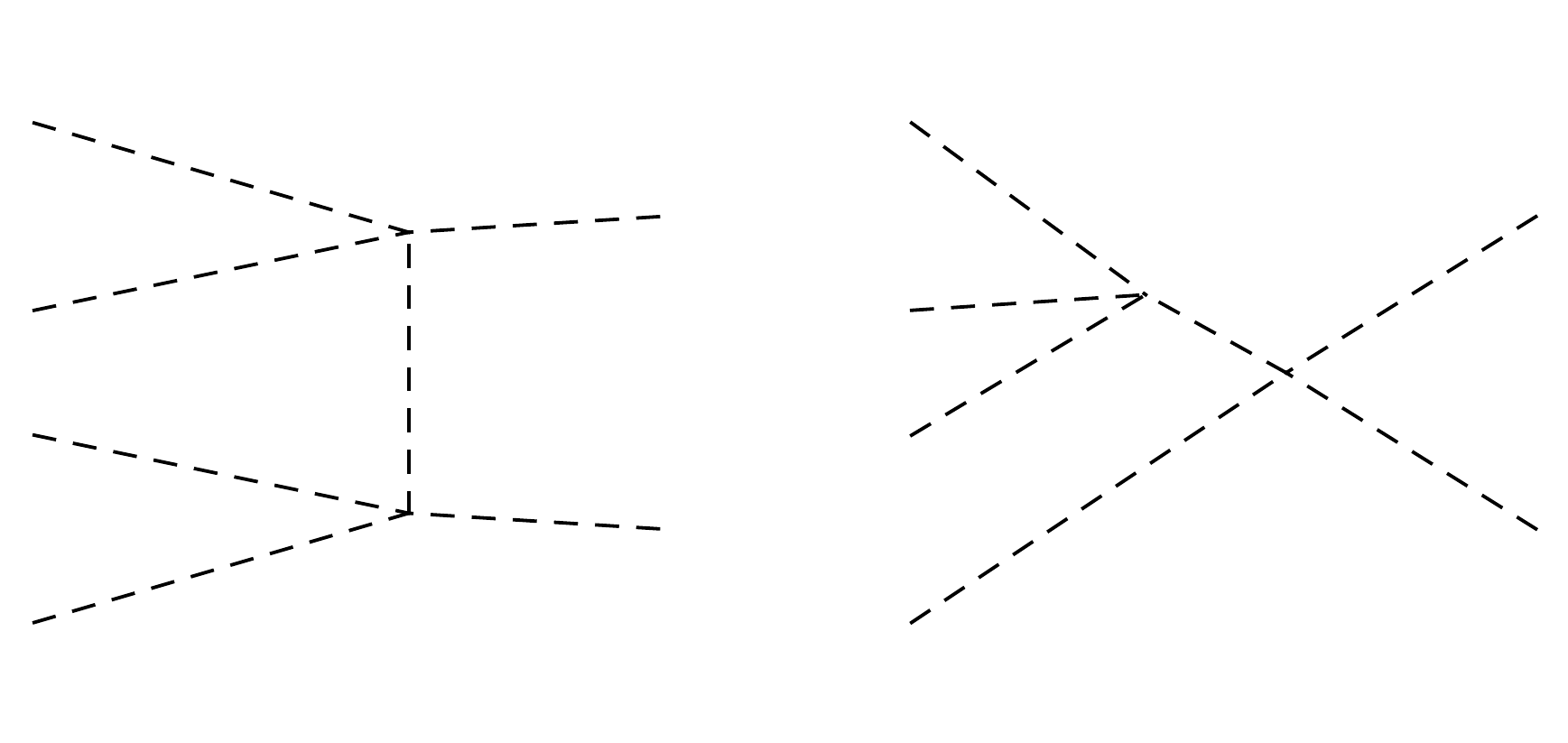}
\caption{Examples of Feynman diagrams for the $4\to 2$ scalar self-annihilation process.}
\label{diagram}
\end{center}
\end{figure}

The DM and visible sectors seize to be in {\it chemical} equilibrium with each other when $\langle\sigma_{ss\to bb}v \rangle n_s/H=1$. However, the $s$ particles can maintain chemical equilibrium among themselves if number-changing interactions (namely, 4-to-2 annihilations with only DM particles both in the initial and final states, see Fig.~\ref{diagram}) are still active. The condition for this so-called cannibalism is given by
\be
\label{cannibalismcondition}
\left. \frac{\langle\sigma_{ss\to b\bar{b}}v \rangle n_s}{\langle\sigma_{4\to 2}v^3 \rangle n_s^3}\right|_{x_{\rm FO}} \simeq \frac{\pi^2}{81\sqrt{3}}\frac{\lambda_{hs}^2}{\lambda_s^4}\,x_{\rm FO}^3\,e^{2x_{\rm FO}} < 1\,,
\ee
where we used 
\begin{equation}
\label{4to2_cross_section}
\langle\sigma_{4\to 2}v^3\rangle \simeq \frac{81\sqrt{3}}{32\pi}\,\frac{\lambda_s^4}{m_s^8} ,
\end{equation}
in the non-relativistic approximation~\cite{Tenkanen:2016jic}, and where $x_{\rm FO}$ is given by Eq.~\eqref{FO condition}.
In this case, the DM abundance is driven by the 4-to-2 annihilations and not anymore by the subdominant annihilations into SM particles. The Boltzmann equation governing the DM number density, Eq.~\eqref{Boltz eq general}, becomes
\begin{equation}
\label{Boltzmann_FO_cannibalism}
\frac{dn_{s}}{dt}+3\,H\,n_{s}=-\left\langle \sigma_{4\to 2} v^3\right\rangle \left[n_{s}^{4}-n_s^2\left(n_{s}^\text{eq}\right)^{2}\right].
\end{equation}

If Eq.~\eqref{cannibalismcondition} was satisfied, the DM freeze-out is given by the decoupling of the 4-to-2 annihilations, defined by
\be
\left.\frac{\langle\sigma_{4\to 2}v^3 \rangle n_s^3}{H}\right|_{T=T_\text{FO}^\text{c}}  = 1\,,
\ee
as can be inferred from Eq. \eqref{Boltzmann_FO_cannibalism}. The time of freeze-out then is
\be
\label{xcFO}
x^{\rm c}_{\rm FO} \equiv\frac{m_s}{T_\text{FO}^\text{c}}= W\left[0.2\,\lambda_s^{4/3}\left(\frac{H_{\rm EW}}{m_h}\right)^{-1/3}\left(\frac{m_s}{\rm GeV}\right)^{-1/6} \right]\, ,
\ee
where $W=W[\lambda_s,\,m_s,\,H_\text{EW}]$ is the 0-branch of the Lambert $W$ function. The DM abundance then becomes (see again Appendix~\ref{appendix})
\begin{align}\label{nsFO}
	\frac{\Omega_{s}\,h^{2}}{0.12}\simeq & \,3\times 10^{-7}\, (x^{\rm c}_{\rm FO})^{3/2}\,e^{-x^{\rm c}_{\rm FO}}\nonumber\\
	&\times \left(\frac{T_{\rm end}}{m_h}\right)^{3/4}\left(\frac{H_{\rm EW}/m_h}{10^{-16}}\right)^{-3/2}\,\left(\frac{m_s}{\rm GeV}\right)\,.
\end{align}
When cannibalism is active, the 4-to-2 annihilations tend to increase the DM temperature with respect to the one of the SM bath~\cite{Carlson:1992fn}.
However, we have checked that in all cases the DM and SM particles were still in {\it kinetic} equilibrium at the time of DM freeze-out, so that temperature of the $s$ particle heat bath was the same as the SM temperature $T$. The condition for this is $\left. \langle\sigma_{sb\to sb}v \rangle n_b/H\right|_{x^{\rm c}_{\rm FO}} > 1$, where we have taken for simplicity $\langle\sigma_{sb\to sb}v \rangle \simeq \langle\sigma_{ss\to b\bar{b}}v \rangle$ and $n_b$ is the $b$-quark number density. 

\begin{figure}
\begin{center}
\includegraphics[width=.49\textwidth]{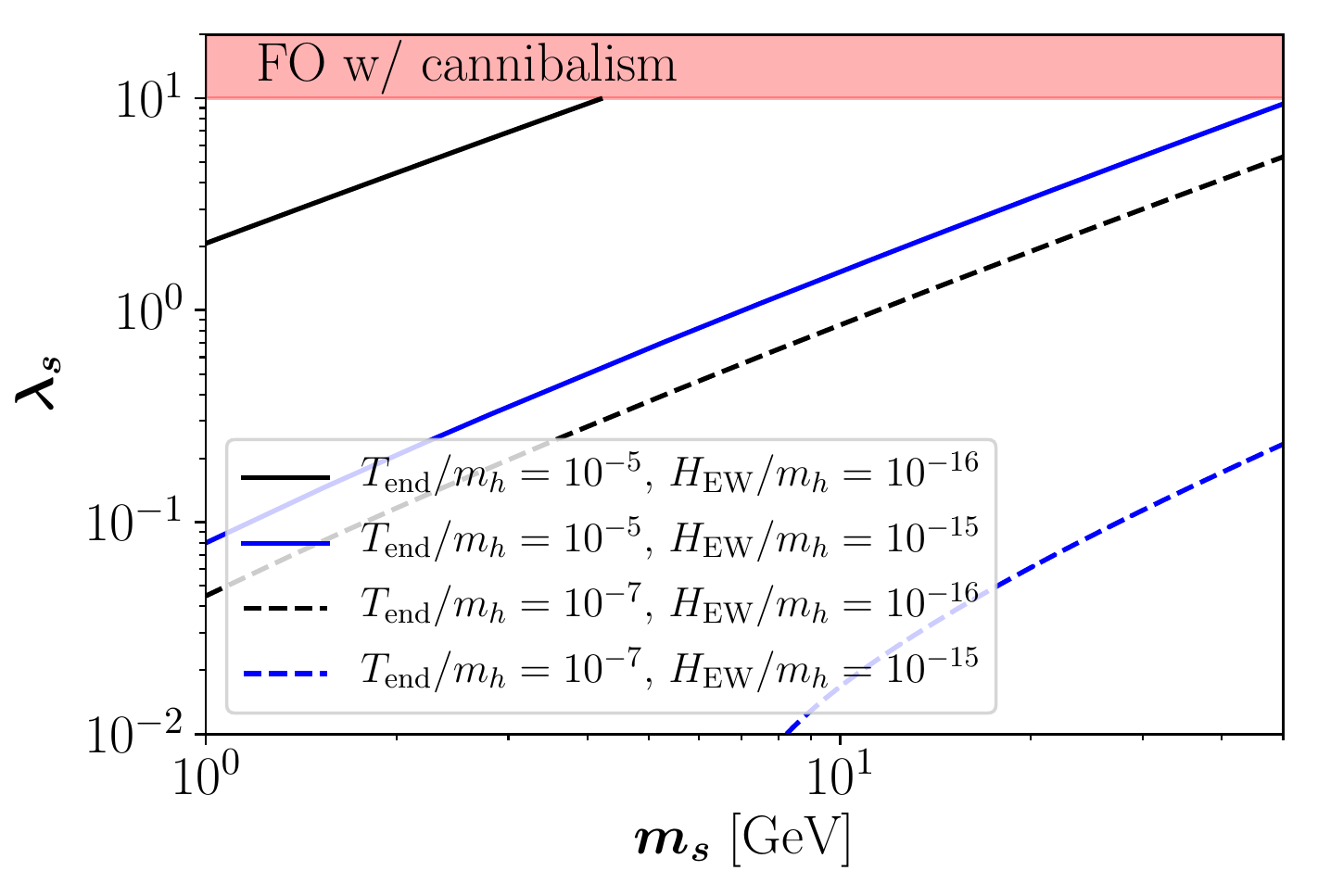}
\caption{DM freeze-out with cannibalism. Parameter space giving rise to the observed DM relic abundance, for $\lambda_{hs}=10^{-3}$. The red region corresponds to $\lambda_s>10$.}
\label{FOwithCannibalSlices}
\end{center}
\end{figure}
Similar to Fig.~\ref{FOwithoutCannibalSlices}, Fig.~\ref{FOwithCannibalSlices} also shows slices of the parameter space that give rise to the observed DM relic abundance.
Here the cosmological parameters are fixed, $H_\text{EW}/m_h=10^{-16}$ (black lines) and $10^{-15}$ (blue lines), and $T_\text{end}/m_h=10^{-7}$ (dashed lines) and $10^{-5}$ (solid lines), while we scan over the particle physics parameters $\lambda_s$ and $m_s$ for a fixed $\lambda_{hs}=10^{-3}$.
The upper band in red, corresponding to $\lambda_s>10$, is not considered.
As in the previous case without cannibalism, an increase in the dilution factor has to be compensated with a higher DM abundance at the freeze-out.
In this case with cannibalism, this requires a smaller annihilation cross-section and hence a small $\lambda_s$ or a heavier DM.
The behavior with respect to $\lambda_{hs}$  and the cosmological parameters is very similar to the case without cannibalism (see Fig.~\ref{FOwithoutCannibalSlices}) and is therefore not presented in this figure.
\\

\begin{figure*}
\begin{center}
\hspace{8.2cm}
\includegraphics[width=.46\textwidth]{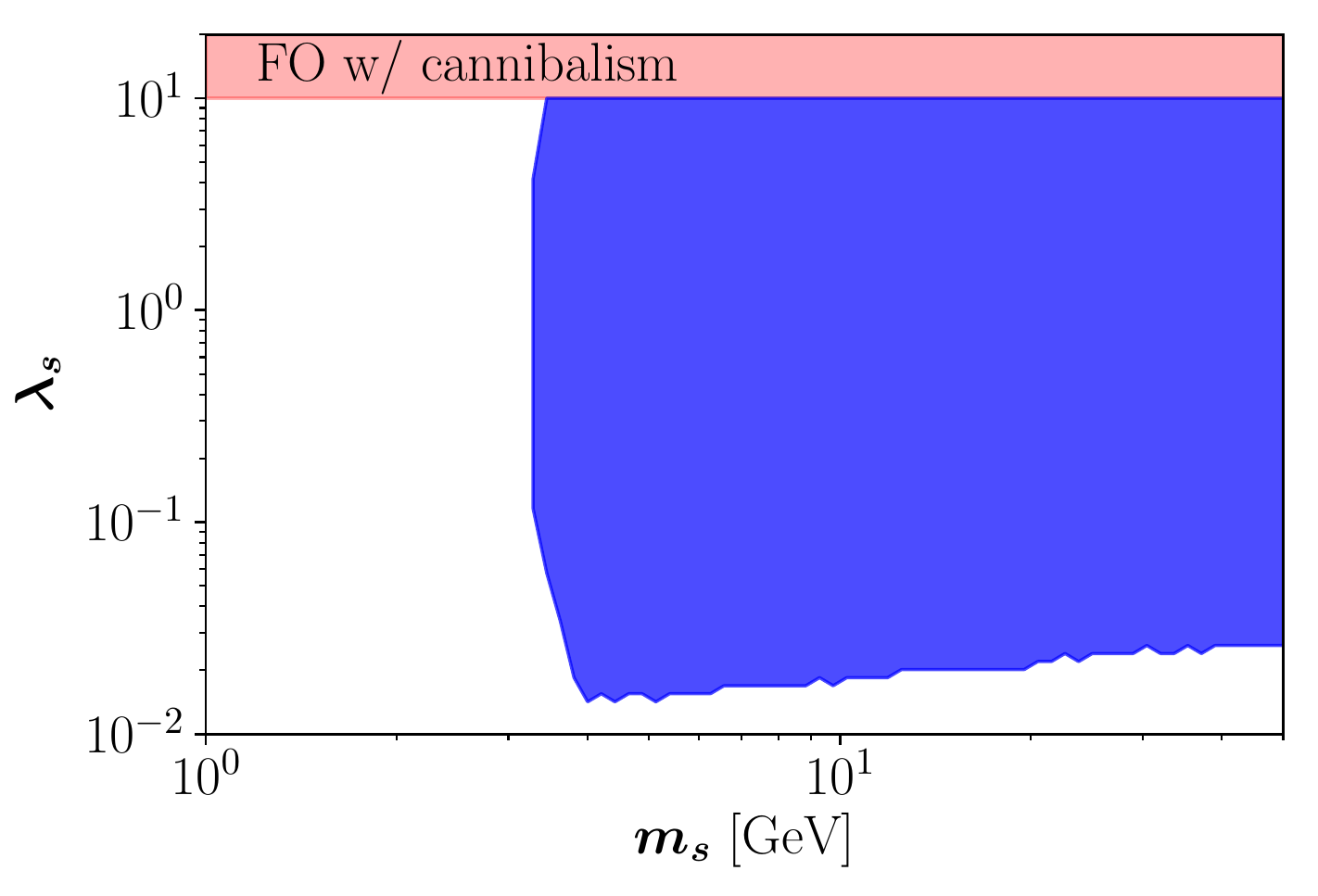}\\
\includegraphics[width=.46\textwidth]{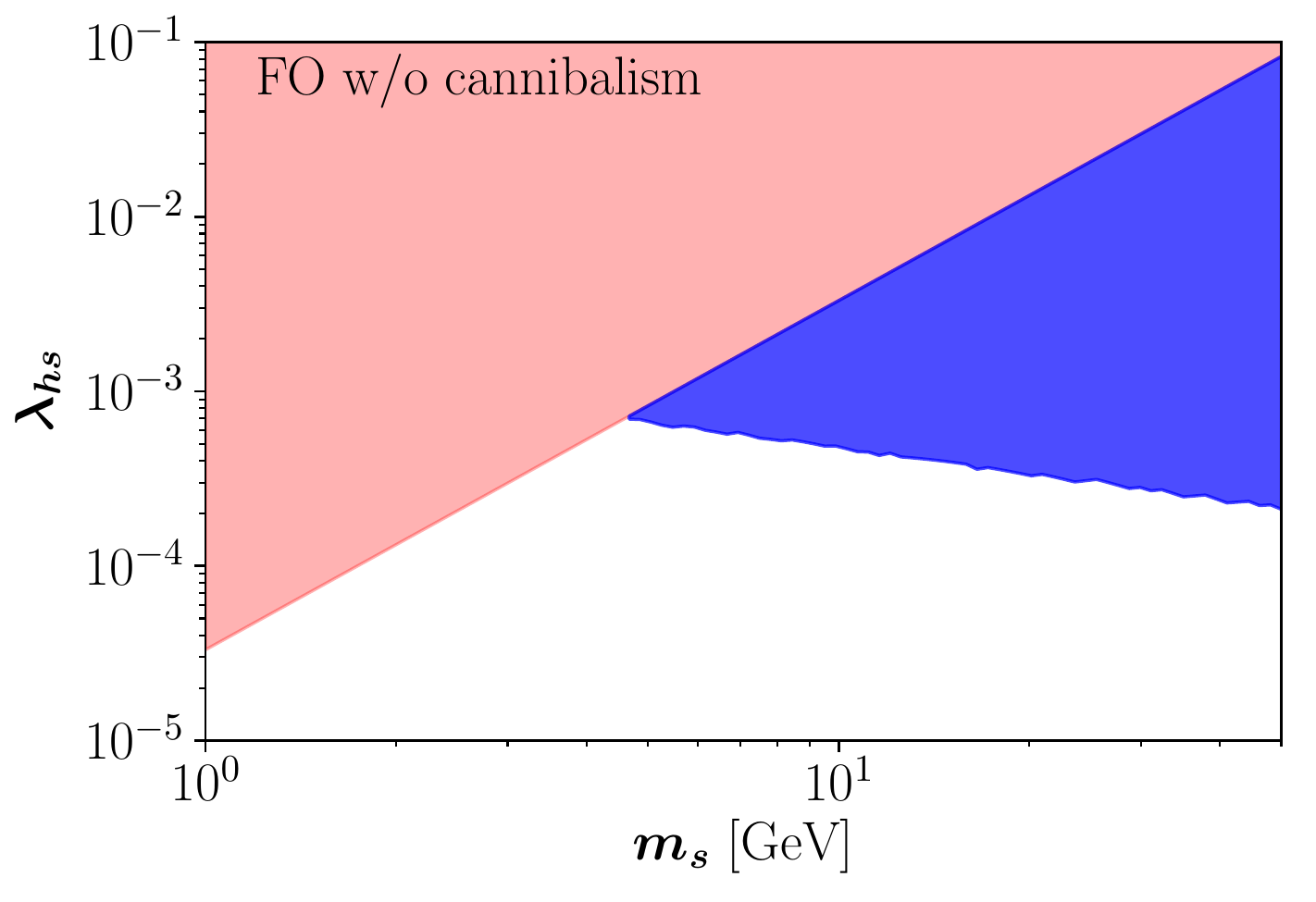}
\includegraphics[width=.46\textwidth]{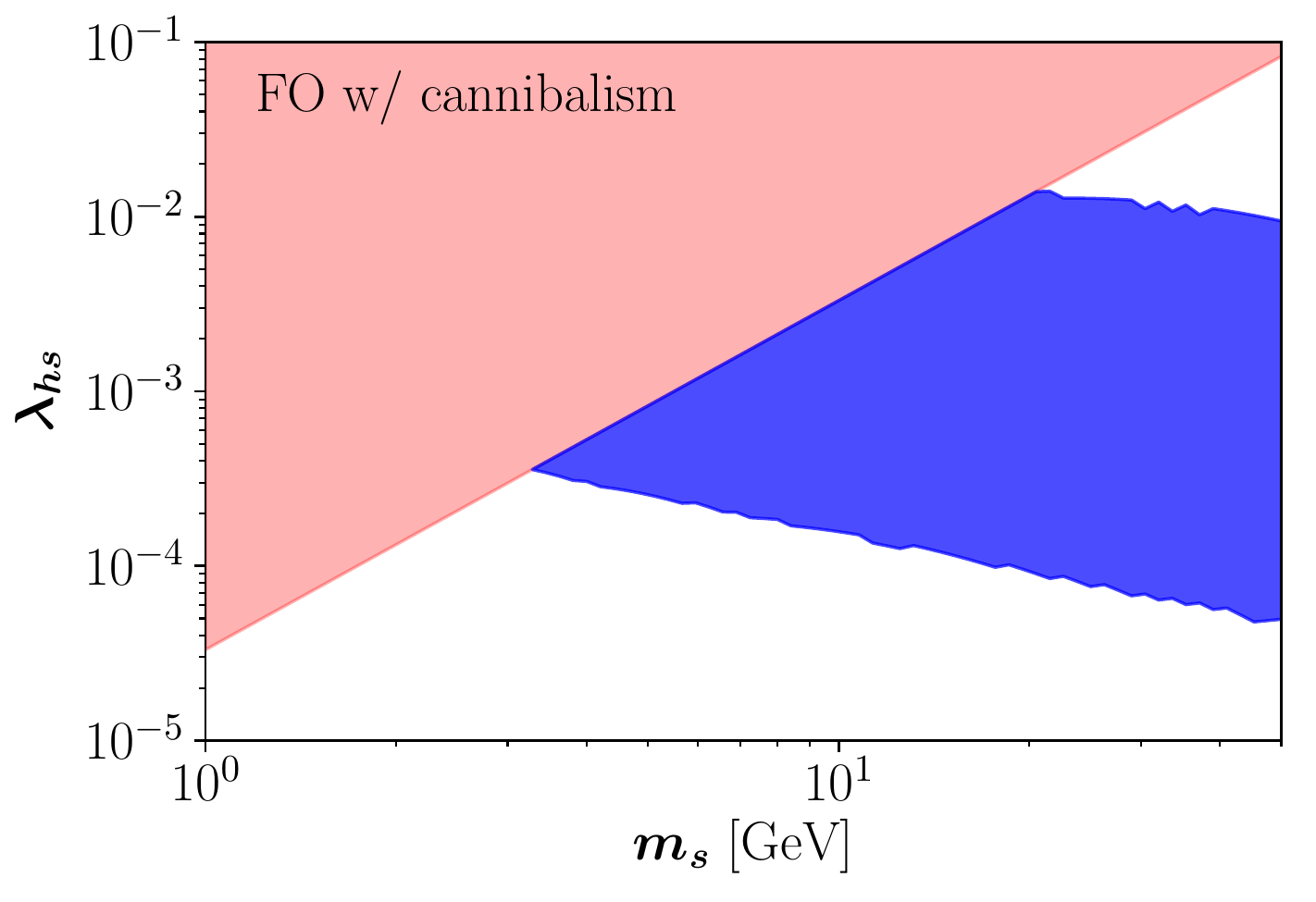}
\includegraphics[width=.46\textwidth]{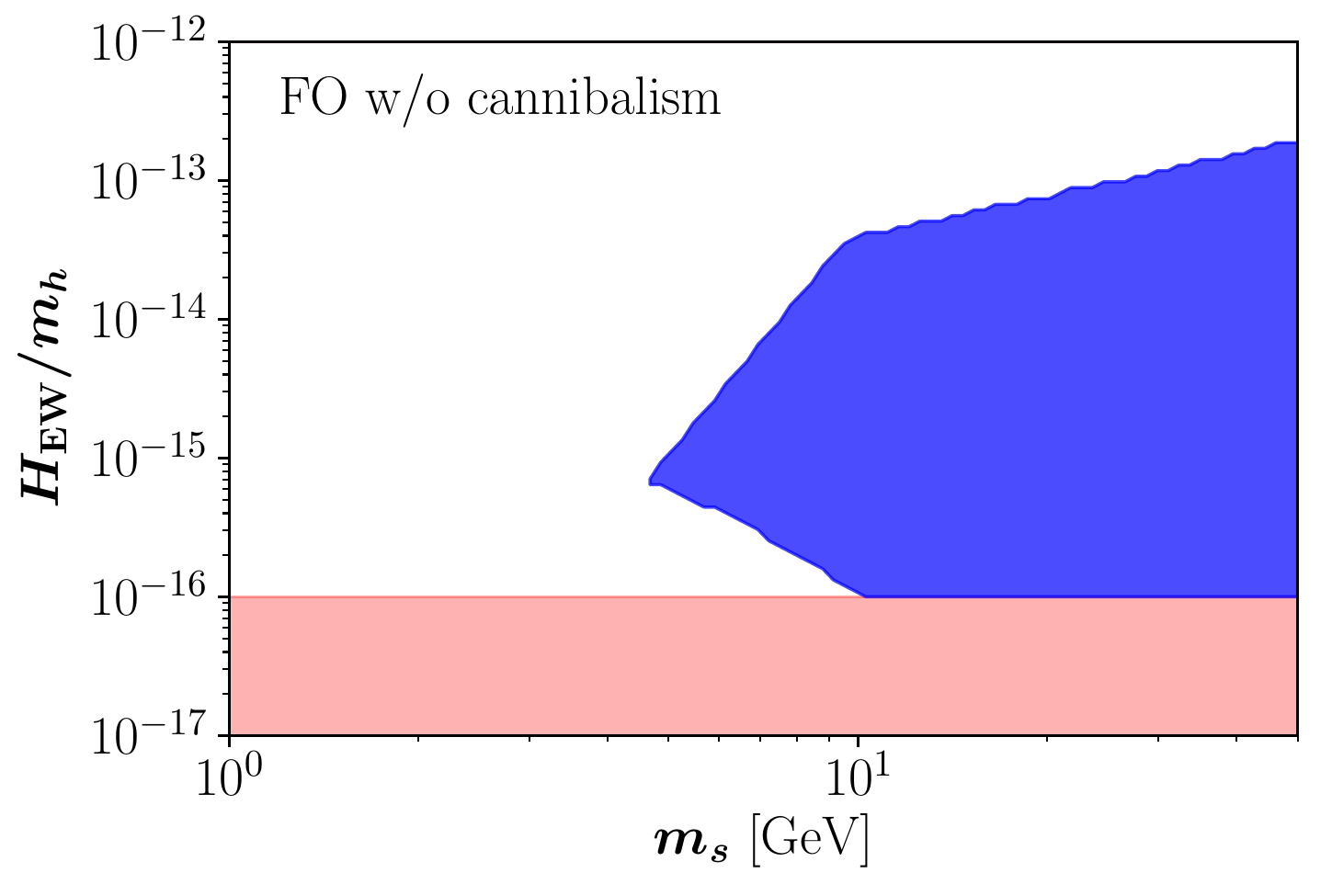}
\includegraphics[width=.46\textwidth]{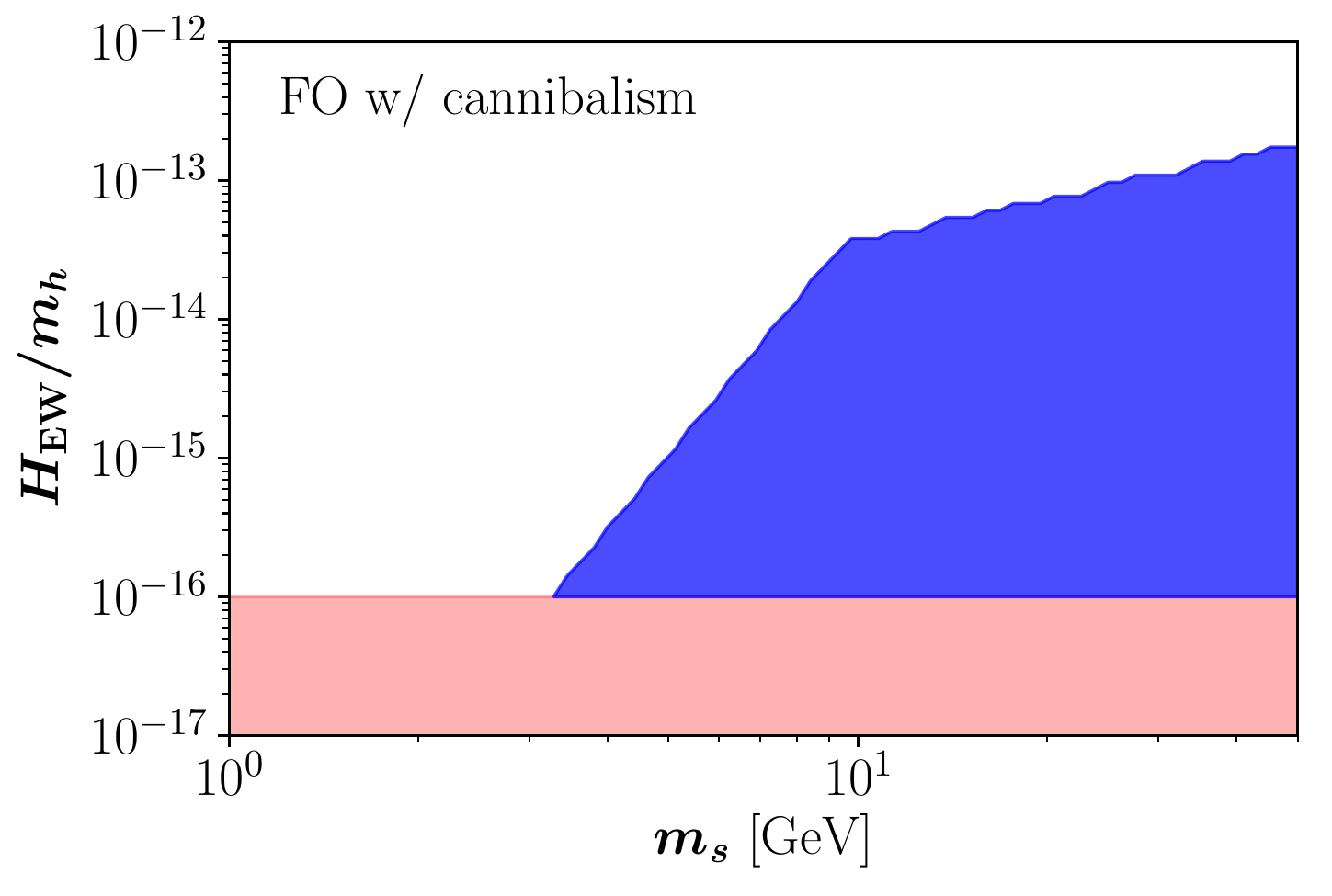}
\includegraphics[width=.46\textwidth]{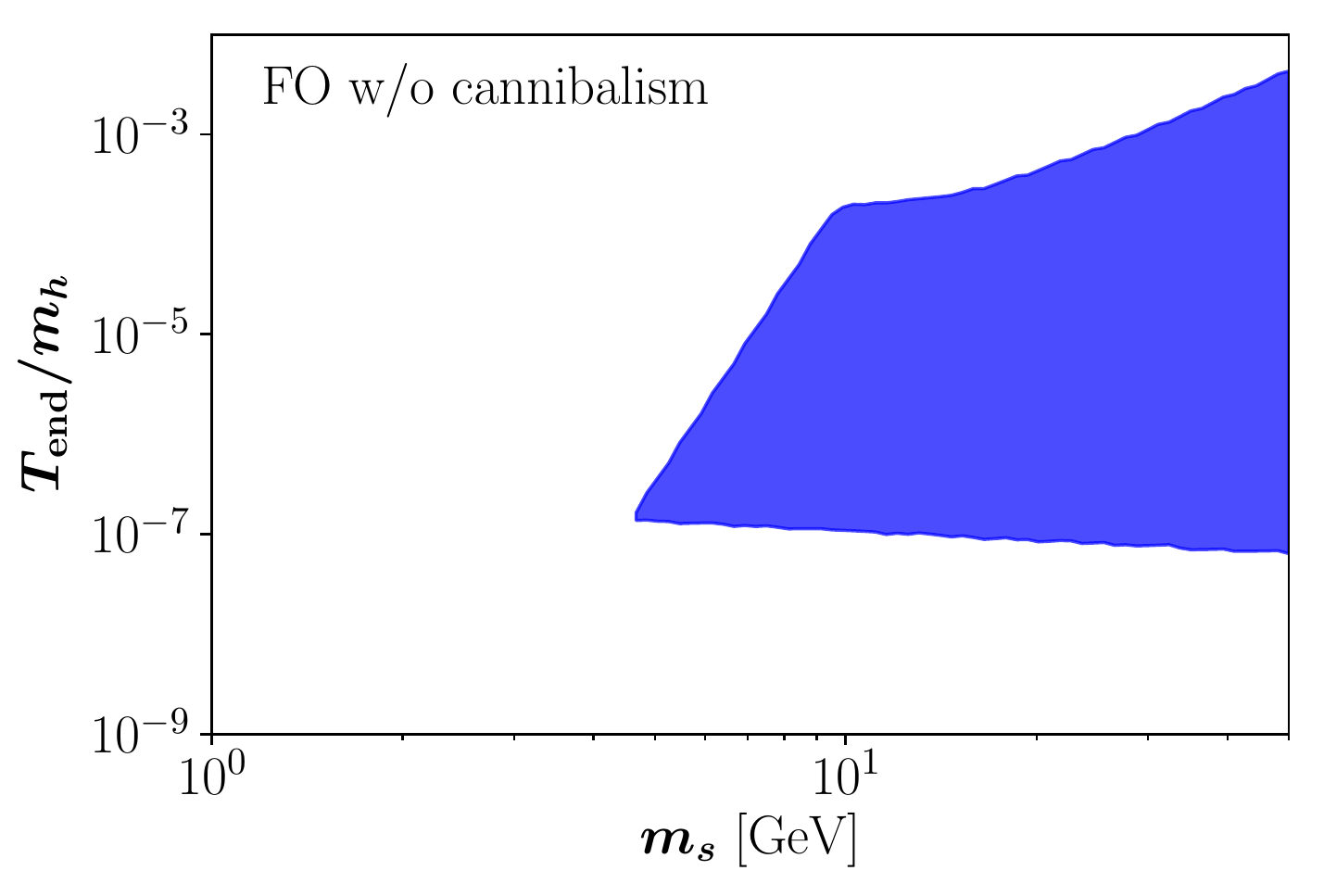}
\includegraphics[width=.46\textwidth]{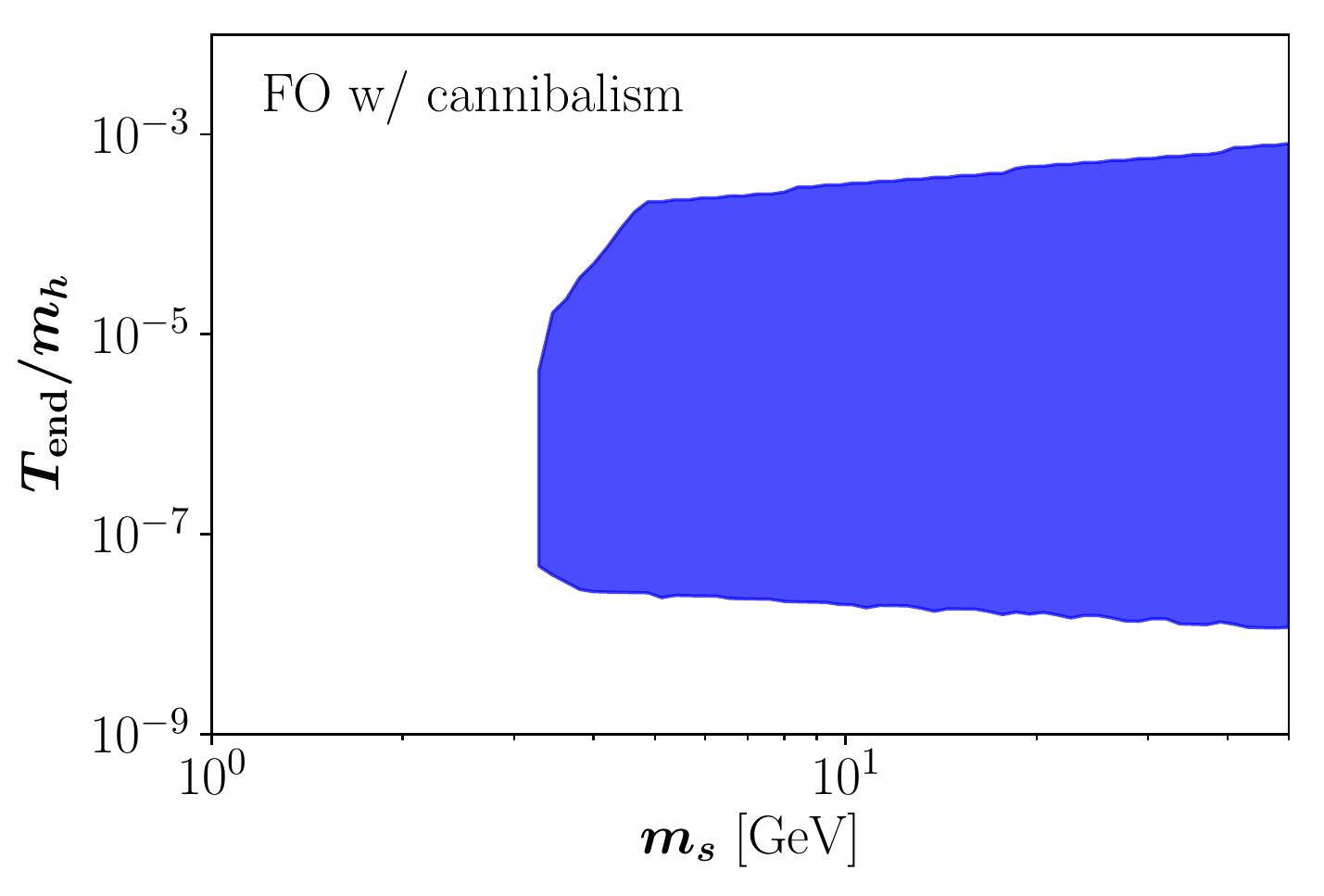}
\caption{DM freeze-out without (left column) and with (right column) cannibalism. Parameter space giving rise to the observed DM relic abundance. The red regions correspond to the constraints discussed in Section~\ref{sec:constraints}: the SM temperature after the matter-like component has decayed into SM particles must be larger than the BBN temperature and small enough not to not re-trigger DM production, Eq. (\ref{tend over mh constr}); the DM freeze-out occurs while the $s$ particles are non-relativistic, $x_{\rm FO}>3$; in a MD Universe $H_{\mathrm{EW}}/m_h >1.76\times 10^{-16}$; the portal coupling has to satisfy $\lambda_{hs}<2\,m_s^2/v^2$ and $\lambda_{hs}\geq \lambda_{hs}^\text{eq}$ with $\lambda_{hs}^\text{eq}$ given by Eq.~\eqref{portalcoupling}. Other observational constraints are shown in Fig.~\ref{Detection}.}
\label{FO}
\end{center}
\end{figure*}
Before closing this subsection, we present the results of an extensive scan over the parameter space for the DM freeze-out without (left column) and with (right column) cannibalism in Fig.~\ref{FO}.
The blue regions produce the observed DM relic abundance, whereas the red regions correspond to the constraints discussed in Section~\ref{sec:constraints}.
The plots generalize the results of Figs.~\ref{FOwithoutCannibalSlices} and~\ref{FOwithCannibalSlices}.
First, let us note that the usual RD scenario can be recovered by taking $H_\text{EW}/m_h=H_\text{EW}^\text{rad}/m_h\simeq 1.76\times 10^{-16}$ and $T_\text{end}/m_h=1$.
This corresponds to $\lambda_{hs}\simeq 10^{-1}$, in the case where DM mainly annihilates into $b$-quarks ($m_b<m_s\lesssim 50$~GeV) and does not undergo a cannibalism phase.
In the MD scenario the Higgs portal coupling $\lambda_{hs}$ can reach much smaller values down to $\mathcal{O}(10^{-4})$.
Such small values naturally need large dilution factors, characterized by large expansion rates $H_\text{EW}/m_h$ up to $\mathcal{O}(10^{-13})$ and low temperatures for the end of the MD era, $T_\text{end}/m_h$ down to $\mathcal{O}(10^{-8})$. In the case with cannibalism, $\lambda_{hs}\lesssim 10^{-2}$ while $\lambda_s\gtrsim 10^{-2}$ due to the fact that the DM annihilation into SM particles must decouple earlier than the 4-to-2 annihilations. Finally, we note that in the scenario where freeze-out occurs during a standard RD phase, cannibalism would generically require non-perturbative values of $\lambda_s$. As shown above, in the MD case the detailed effect of non-vanishing self-interactions can easily be taken into account, as the required values for $\lambda_s$ can be much smaller. This result, along with its observational consequences that we will present in Section~\ref{constraints}, are among the most important novelties of this work.

\subsection{The Freeze-in Case}
\label{sec:freezein}

In this subsection we assume, for simplicity, the mass hierarchy $m_s < m_h/2$, as we take the Higgs decay into two $s$ to be the dominant production mechanism for DM. A more general analysis  is again presented in Ref.~\cite{Bernal:2018kcw} for the pure freeze-in case without cannibalism.

\subsubsection{Freeze-in without Cannibalism}
\label{sec:FI w/o cann}

The DM number density can again be computed using the Boltzmann equation~\eqref{Boltz eq general}, which in the absence of DM self-interactions is
\begin{equation}
\frac{dn_s}{dt}+3\,H\,n_s=2\,\frac{K_{1}(\frac{m_h}{T})}{K_{2}(\frac{m_h}{T})}\,\Gamma_{h\rightarrow ss}\,n_h^{{\rm eq}}\,,\label{sBoltzmann}
\end{equation}
where $\Gamma_{h\rightarrow ss}$ is the partial decay width of the Higgs into two $s$-particles and $n_h^\text{eq}$ is its equilibrium number density. These quantities are given by
\be
\label{Gamma}
\Gamma_{h\rightarrow ss} 
= \frac{\lambda_{hs}^2\,m_h}{64\pi \lambda_h}\sqrt{1-\left(\frac{2m_s}{m_h}\right)^2} \,,
\ee
\be
\label{nheq}
n_h^{\rm eq}(T) = \left(\frac{m_h\,T}{2\pi} \right)^{3/2} e^{-\frac{m_h}{T}}\,.
\ee
By then performing a change of variables, $\chi_s = n_s\,a^3$, where $\chi_s$ is the comoving $s$ number density and $a$ is the scale factor, we get the comoving DM number density at infinity\footnote{Assuming that the initial DM abundance vanishes. For extended discussion on the validity of this assumption, see Refs.~\cite{Dev:2013yza,Nurmi:2015ema,Kainulainen:2016vzv}.}
\bea
\label{chi_infty}
\chi_s^\infty &=& 2\, \Gamma_{h\rightarrow ss}\int_0^\infty {\rm dln}a\left(\frac{m_h\,T}{2\pi}\right)^{3/2}e^{-m_h/T}\frac{a^3}{H(a)}\frac{K_1(\frac{m_h}{T})}{K_2(\frac{m_h}{T})} \nn\\ 
&\simeq& 6.3\,\frac{\Gamma_{h\rightarrow ss}}{H_\text{EW}}\,n_h^{\rm eq}(m_h)\,,
\eea
where we have normalized the scale factor so that $a(T=m_h)\equiv a_{\rm EW}=1$. The numerical value of the above integral is not sensitive to the upper limit of integration, and we have set it for convenience to $a\to \infty$. As shown in the Appendix \ref{appendix}, the DM abundance today can then be expressed as
\begin{align}
\label{abundance_notherm}
	\frac{\Omega_{s}\,h^{2}}{0.12}\simeq &\, 2\times10^{22}\,g_*(m_h)^{-1/4}\,\lambda_{hs}^2\nonumber\\
	& \times\left(\frac{H_{\mathrm{EW}}/m_h}{10^{-16}}\right)^{-5/2}\,\left(\frac{T_{\rm end}}{m_h}\right)^{3/4}\left(\frac{m_{s}}{\mathrm{GeV}}\right)\,,
\end{align}
where we assumed $m_s\ll m_h/2$.

Let us emphasize that the result in Eq.~\eqref{abundance_notherm} only applies to a scenario where the Universe was effectively MD during the DM yield, and therefore it is not, as such, applicable to other scenarios. To retain the usual RD case, one must set $T_{\rm end}=m_h$, use the result of Eq.~\eqref{HEW} for $H_{\rm EW}$, and use the newly calculated prefactor $11.4$ in Eq.~\eqref{chi_infty} instead of $6.3$ which we obtained above. These account for the facts that in our case not only there was entropy production at the end of the early MD phase but also that the expansion rate of the Universe at the time of DM freeze-in was different from that in the usual RD case.

\begin{figure}
\begin{center}
\includegraphics[width=.49\textwidth]{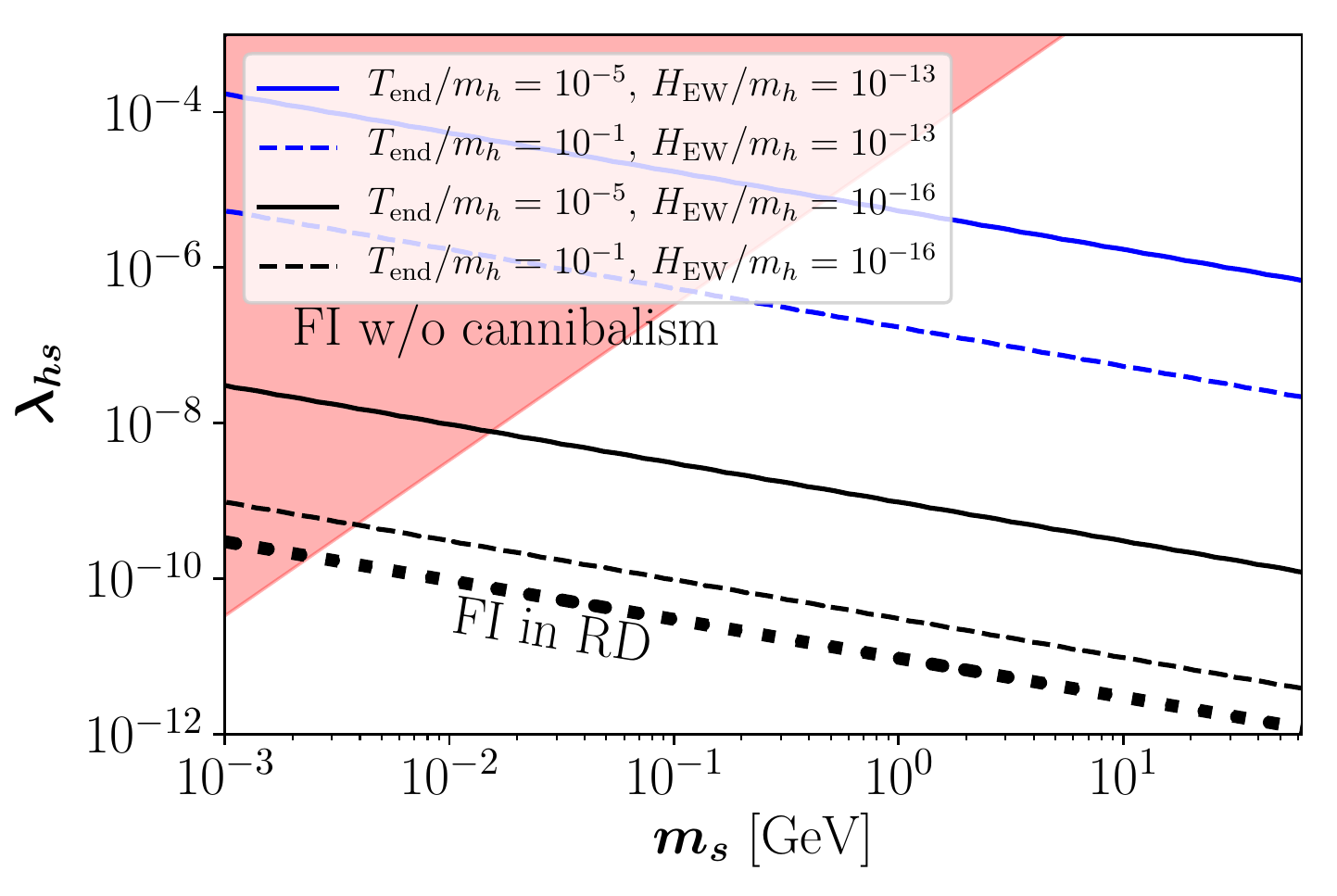}
\includegraphics[width=.49\textwidth]{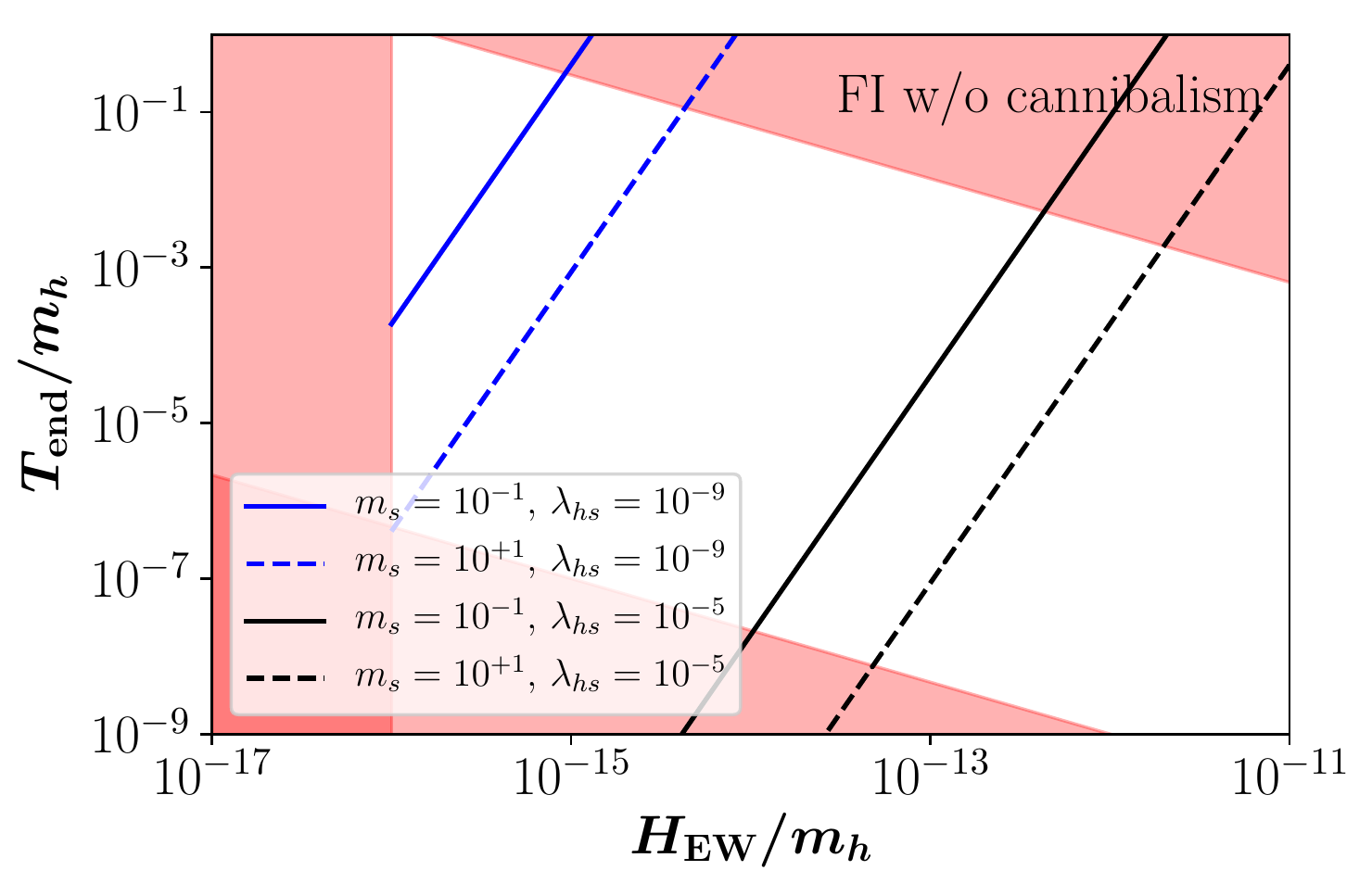}
\caption{DM freeze-in without cannibalism. Parameter space giving rise to the observed DM relic abundance.
The black dotted line shows the parameters yielding the correct DM abundance in the usual RD scenario.
The red regions correspond to the constraints discussed in Section~\ref{sec:constraints}. Other observational constraints are shown in Fig.~\ref{Detection}.}
\label{FIwithoutCannibalSlices}
\end{center}
\end{figure}
Fig.~\ref{FIwithoutCannibalSlices} shows slices of the parameter space that give rise to the observed DM relic abundance.
On the upper panel the cosmological parameters are fixed, $H_\text{EW}/m_h=10^{-16}$ (black lines) and $10^{-13}$ (blue lines), and $T_\text{end}/m_h=10^{-5}$ (solid lines) and $10^{-1}$ (dashed lines), while we scan over the relevant particle physics parameters ($\lambda_{hs}$ and $m_s$).
The upper left corner in red, corresponding to, $\lambda_{hs}>2\,m_s^2/v^2$, is excluded.
The figure shows again that an increase in the dilution factor due to either an enhancement of the Hubble expansion rate $H_\text{EW}$ or a decrease in the temperature $T_\text{end}$ when the MD era ends has to be compensated with a higher DM abundance at the freeze-out. This requires an increase in either $m_s$ or the DM production via the Higgs decay (i.e. a bigger $\lambda_{hs}$).
The thick dotted black line corresponds to the DM production in the usual RD scenario, characterized by $T_\text{end}/m_h=1$ and $H_\text{EW}/m_h=10^{-16}$.
We note that, as expected, in the MD scenario the values for the required values for Higgs portal are always higher than in the RD case.

The same conclusion can be drawn from the lower panel of Fig.~\ref{FIwithoutCannibalSlices}, where the particle physics parameters are fixed, $m_s=0.1$~GeV (solid lines) and $10$~GeV (dashed lines), and $\lambda_{hs}=10^{-9}$ (blue lines) and $10^{-5}$ (black lines), while we scan over the cosmological parameters.
The left band corresponds to a scenario which is not MD \linebreak ($H_\text{EW}/m_h<10^{-16}$). The lower left and the upper right corners correspond to scenarios where the resulting SM temperature after the MD era ends is either too small for successful BBN or so large that it re-triggers the DM yield, respectively. All three cases are excluded from our analysis. Observational constraints on the scenario will be discussed in Section~\ref{constraints}.

As in the case of freeze-out, the result of Eq.~\eqref{abundance_notherm} is the final DM abundance only if number-changing DM self-interactions do not become active and the $s$ particles do not reach chemical equilibrium with themselves. This is the scenario we will now turn into.

\subsubsection{Freeze-in with Cannibalism}
\label{sec:cannibalism}

Let us now calculate the final DM abundance following the thermalization and consequent cannibalism phase of the $s$ particles. In this case, the Boltzmann equation~\eqref{Boltz eq general} is
\begin{align}
	\frac{dn_s}{dt}+3\,H\,n_s= & \, 2\,\frac{K_{1}(\frac{m_h}{T})}{K_{2}(\frac{m_h}{T})}\,\Gamma_{h\rightarrow ss}\,n_h^{{\rm eq}}\nonumber\\
	&  - \left\langle \sigma_{4\to 2} v^3\right\rangle \left[n_{s}^{4}-n_s^2\left(n_{s}^\text{eq}\right)^{2}\right]\,,
\label{sBoltzmann_FI_cannibalism}
\end{align}
where $\Gamma_{h\rightarrow ss}$ and $n_h^{{\rm eq}}$ are again given by Eqs.~\eqref{Gamma} and~\eqref{nheq}, respectively, and $\left\langle \sigma_{4\to 2} v^3\right\rangle$ by Eq.~\eqref{4to2_cross_section}. 

For values of the portal coupling required by non-thermalization of the hidden sector with the SM sector $\lambda_{hs}\lesssim (H_{\rm EW}/m_h)^{1/2}$, Eq.~\eqref{portalcoupling}, the initial $s$ particle number density in the hidden sector produced by Higgs decays is always smaller than the corresponding equilibrium number density. Thus, if the self-interactions are sufficiently strong (see below), the $s$ particles can reach chemical equilibrium with themselves by first increasing their number density via 2-to-4 annihilations, and then undergo cannibalism when they become non-relativistic, as discussed in e.g. Refs.~\cite{Bernal:2015xba,Heikinheimo:2016yds,Heikinheimo:2017ofk}. A possible caveat to this is the case where $m_s$ is close to $m_h$, as then the eventual {\it dark freeze-out} would occur before the yield from the SM sector has ended. In that case, the production mechanism is dubbed as {\it reannihilation}~\cite{Cheung:2010gj,Chu:2011be}. Because in that case the $s$ particles would not, in general, be in thermal equilibrium at the time of their freeze-out, finding the correct DM abundance requires solving the Boltzmann equation for the DM distribution function instead of number density, which is beyond the scope of this work. In this paper we therefore choose an approach where we solve the Boltzmann equation for DM number density but highlight the regime in our results where reannihilations could potentially alter our conclusions, and leave solving the Boltzmann equation for DM distribution function for future work. Because the freeze-in yield has ended by $T\sim 0.1 m_h$~\cite{Chu:2011be}, we take this regime to be determined by $m_s\gtrsim 10$~GeV. As we will show, this is only a small part of the observationally interesting parameter space, especially for DM self-interactions.

In the following, we will solve Eq.~\eqref{sBoltzmann_FI_cannibalism} in the limit where the self-interactions of $s$ are large, to complement the usual freeze-in scenario discussed above. Note that the 2-to-2 scalar self-annihilations do not have a net effect on the final DM abundance and are therefore not included in Eq.~\eqref{sBoltzmann_FI_cannibalism}.

The number-changing $s$ self-interactions in Eq.~\eqref{sBoltzmann_FI_cannibalism} become active if
\be
\label{selfinteractions}
\left. \frac{\langle\sigma_{4\to 2}v^3\rangle \left(n_s^{\rm init}\right)^3}{H}\right|_{a_{\rm nrel}} > 1 \,,
\ee
where $n_s^{\rm init}(a_{\rm nrel}) = \chi_s^\infty (a_\text{EW}/a_{\rm nrel})^3$ is the initial $s$ particle abundance produced by Higgs decays, where $\chi_s^\infty$ is given by Eq. \eqref{chi_infty}, and we have invoked the principle of detailed balance. The scale factor $a_\text{nrel}$ when the $s$ particles become non-relativistic can be solved from
\be
\frac{p_s}{m_s} \simeq \frac{m_h}{2m_s}\frac{a_{\rm EW}}{a_{\rm nrel}} \simeq 1 \,,
\ee
so that $a_{\rm nrel} \simeq m_h/(2m_s)$ (recall that $a_{\rm EW}=1$). Here we assumed $m_s\ll m_h/2$, so that the initial $s$ particle momenta are $p\simeq m_h/2$. As discussed in Refs.~\cite{Heikinheimo:2016yds,Enqvist:2017kzh}, it indeed suffices to evaluate Eq.~\eqref{selfinteractions} at $a_{\rm nrel}$, which is the latest moment when the $s$ particles can reach chemical equilibrium with themselves. 

Reminiscent to the standard WIMP case, the final DM abundance only depends on the time of the freeze-out, and therefore the scenario is not sensitive to when the hidden sector thermalization occurs. Thus, the thermalization condition for the $s$ field's quartic self-interaction strength can be solved from Eq.~\eqref{selfinteractions} to be
\be
\label{lambda_threshold}
\lambda^{\rm FI}_s \simeq 6.6\,\lambda_{hs}^{-3/2}\left(\frac{m_s}{\rm GeV}\right)^{1/8}\frac{H_{\rm EW}}{m_h}\,.
\ee
If $\lambda_s<\lambda^{\rm FI}_s$, the final yield is given by Eq.~\eqref{abundance_notherm}; if not, cannibalism has to be taken into account in solving Eq.~\eqref{sBoltzmann_FI_cannibalism}. Therefore, if $\lambda_s>\lambda^{\rm FI}_s$, the $s$ particles thermalize with themselves and the sector exhibits a cannibal phase before the final freeze-out of DM density from the hidden sector heat bath. The time of the dark freeze-out of $s$ particles can be solved in the standard way from Eq.~\eqref{sBoltzmann_FI_cannibalism} as the time when the 4-to-2 interaction rate equals the Hubble expansion rate
\be
\label{DFO}
\left.\frac{\langle\sigma_{4\to 2}v^3\rangle n_s^3}{H}\right|_{T_s^\text{FO}} = 1\, ,
\ee
where $H$ is given by Eq.~\eqref{Hubble} and
\be
\label{ns}
n_s(T_s) = \left(\frac{m_sT_s}{2\pi}\right)^{\frac{3}{2}}e^{-\frac{m_s}{T_s}} = \frac{m_s^3}{(2\pi)^{3/2}}x_s^{-3/2}e^{-x_s} \,,
\ee
where $T_s$ is the temperature of the hidden sector heat bath which in general is not the same as the SM sector temperature, $T_s\neq T$. Here we also introduced the conventional units $x_s\equiv m_s/T_s$.

The relation between $T_s$ and $T$ can be inferred from entropy conservation, as after the thermalization within the hidden sector the two entropy densities are separately conserved. First, consider the times when the $s$ particles are still relativistic, whence
\begin{align}
\label{rhoSMrhoS}
	\zeta & \equiv \left. \frac{\mathfrak{s}_{\rm rad}}{\mathfrak{s}_{\rm hid}}\right|_{\rm rel} = \frac{g_{*\mathfrak{s}}\,T^3}{T_s^3}= g_{*\mathfrak{s}}\left(\frac{\rho_{\rm SM}}{g_*\,\rho_s}\right)^{3/4}\nonumber\\
	& = g_{*\mathfrak{s}}\left(\frac{\rho_{\rm SM}}{g_*(m_h/2)\,n_s^{\rm init}}\right)^{3/4}\,,
\end{align}
where $\mathfrak{s}_{\rm rad}$ and $\mathfrak{s}_{\rm hid}$ are the SM and hidden sector entropy densities, respectively, and $g_{*\mathfrak{s}}$ corresponds to the relativistic degrees of freedom that contribute to the SM entropy density. On the other hand, between the moment when the $s$ particles became non-relativistic and their final freeze-out, the ratio $\zeta$ is
\be
\label{nonrelS}
\zeta = \left. \frac{\mathfrak{s}_{\rm rad}}{\mathfrak{s}_{\rm hid}}\right|_{\rm nrel}
= \frac{2\pi^2(2\pi)^{3/2}\,g_*(T)}{45}\,\frac{T^3}{m_s^3}\,x_s^{1/2}\,e^{x_s} \,,
\ee
where we used $\mathfrak{s}_{\rm hid} = m_s\,n_s(T_s)/T_s$. By equating Eqs.~\eqref{rhoSMrhoS} and~\eqref{nonrelS}, one can express the SM sector temperature $T$ as a function of the hidden sector temperature
\be
\label{SMtemp}
T\simeq 1.7\,\lambda_{hs}^{-1/2}\left(\frac{H_\text{EW}}{m_h}\right)^{1/4}x_s^{-1/6}\,e^{-x_s/3}\,m_s \,.
\ee

The moment of the dark freeze-out can then be calculated be using Eqs.~\eqref{DFO}, \eqref{ns}, \eqref{Hubble} and \eqref{SMtemp}, which give
\be
\label{xFO}
x_s^{\rm FO} = \frac{17}{10} W\left[0.1\,\lambda_s^{16/17}\lambda_{hs}^{3/17}\left(\frac{m_h}{m_s}\right)^{2/17}\left(\frac{H_\text{EW}}{m_h}\right)^{-11/34} \right] ,
\ee
where $W=W[\lambda_s,\,\lambda_{hs},\,m_s,\,H_\text{EW}]$ is again the 0-branch of the Lambert $W$ function.  The final DM abundance after the freeze-out then is

\be
\label{ns_final}
n_s^{\rm final} = \frac{m_s^3}{(2\pi)^{3/2}}(x_s^{\rm FO})^{-3/2}e^{-x_s^{\rm FO}} \,,
\ee 
from which the DM abundance today can be calculated to be
\begin{align}
	\frac{\Omega_{s}\,h^2}{0.12}\simeq & \,3\times10^{8}\,g_{*}(T_{\rm FO})^{-1/4}\nonumber\\
	& \times\left(\frac{n_{s}^{\mathrm{final}}}{T_{\rm FO}^{3}}\right)\,\left(\frac{H_{\rm EW}/m_h}{10^{-16}}\right)^{-3/2}\,\left(\frac{T_{\mathrm{end}}}{m_h}\right)^{3/4}\,\left(\frac{m_{s}}{\mathrm{GeV}}\right) ,\label{ns final}
\end{align}
as shown in the Appendix~\ref{appendix}. Using then Eqs.~\eqref{SMtemp} and~\eqref{ns_final}, we get a relation for $x_{s}^{\mathrm{FO}}$ that takes into account the present DM abundance 
\begin{align}
	x_{s}^{\mathrm{FO}}\simeq&\, 4\times10^{18}\,g_*^{-1/4}\lambda_{hs}^{3/2}\nonumber\\
	&\times\left(\frac{\Omega_s h^2}{0.12}\right)^{-1}\left(\frac{H_{\rm EW}/m_h}{10^{-16}}\right)^{-9/4}\left(\frac{T_{\rm end}}{m_h}\right)^{3/4}\left(\frac{m_{s}}{\mathrm{GeV}}\right).\label{xfo abundance}
\end{align}
Equating this result with Eq.~\eqref{xFO} then gives the connection between the model parameters $m_{s}$, $\lambda_{s}$, $\lambda_{hs}$, $T_{\mathrm{end}}$, $H_{\mathrm{EW}}$ that yields the correct DM abundance.

\begin{figure}
\begin{center}
\includegraphics[width=.49\textwidth]{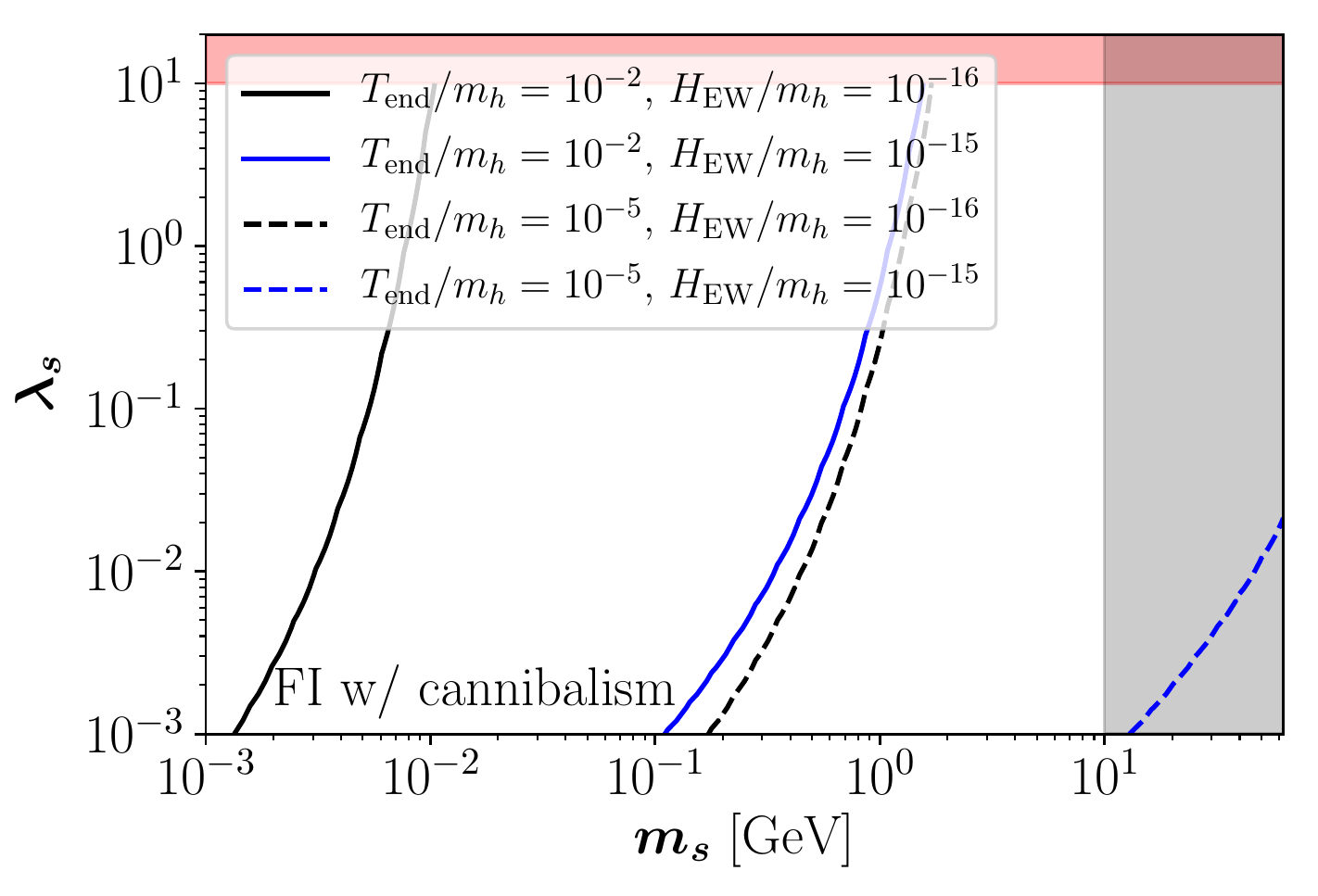}
\includegraphics[width=.49\textwidth]{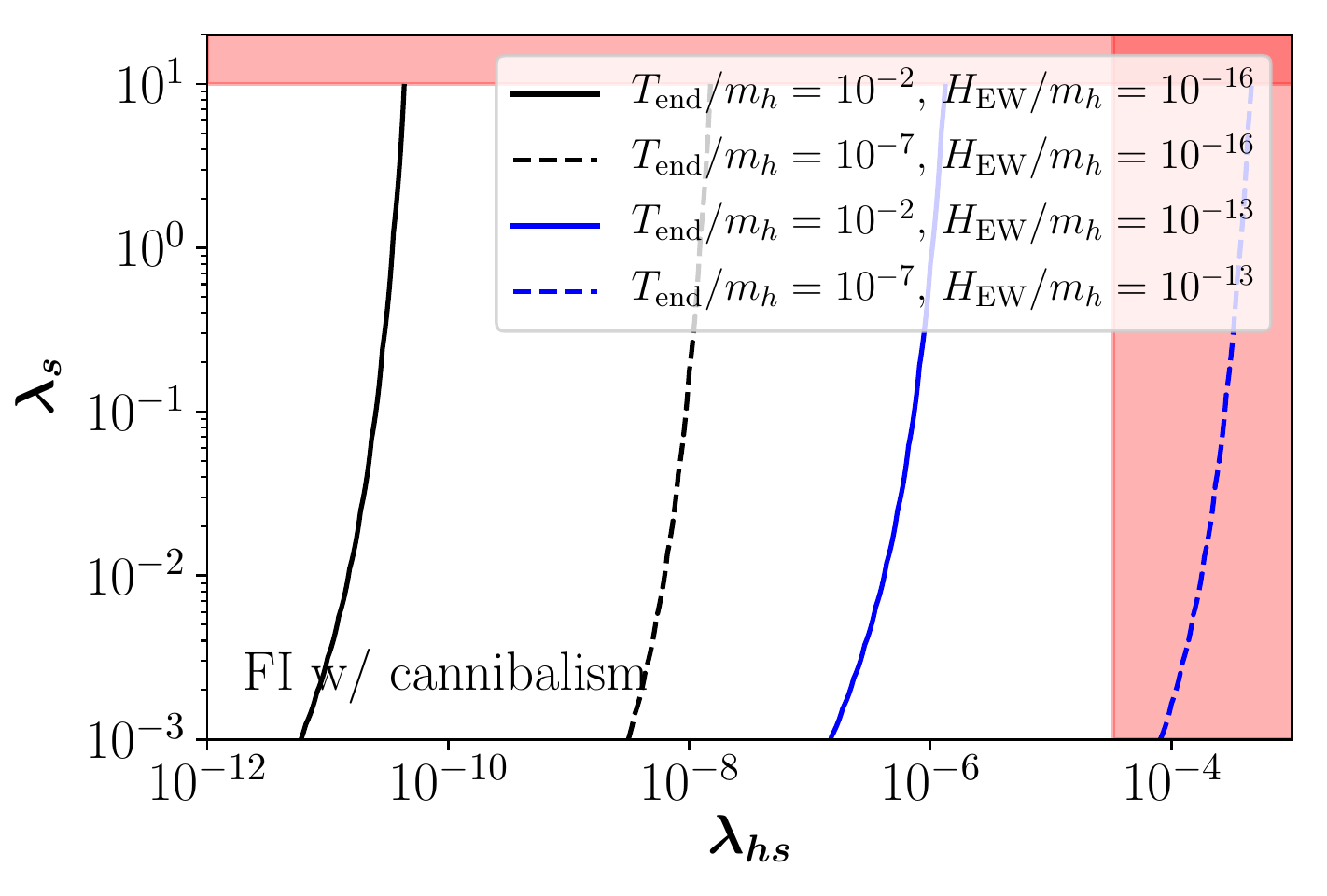}
	\caption{DM freeze-in with cannibalism. Parameter space giving rise to the observed DM relic abundance, for $\lambda_{hs}=10^{-9}$ (upper panel) and $m_s=1$~GeV (lower panel). The red regions correspond to the constraints discussed in Section~\ref{sec:constraints}, and the shaded region in the upper panel to the reannihilation regime. Other observational constraints are shown in Fig.~\ref{Detection}.}
\label{FIwithCannibalSlices}
\end{center}
\end{figure}

Fig.~\ref{FIwithCannibalSlices} shows again slices of the parameter space that give rise to the observed DM relic abundance.
The cosmological parameters are fixed and we scan over the particle physics parameters, fixing $\lambda_{hs}=10^{-9}$ in the upper panel and $m_s=1$~GeV in the lower panel.
The red bands, corresponding to $\lambda_s>10$ (perturbativity bound) and $\lambda_{hs}\gtrsim3\times 10^{-5}$ ($\lambda_{hs} < 2m_s^2/v^2$ in order to avoid a spontaneous symmetry breaking in the $s$ direction) are excluded.
Again, an increase in the dilution factor due to either an enhancement of the Hubble expansion rate $H_\text{EW}$ or a decrease in the temperature $T_\text{end}$ when the MD era ends has to be compensated with a higher DM abundance at the dark freeze-out. This requires a smaller 4-to-2 annihilation cross-section and hence a small $\lambda_s$.
\\

\begin{figure*}
\begin{center}
\hspace{8.2cm}
\includegraphics[width=.46\textwidth]{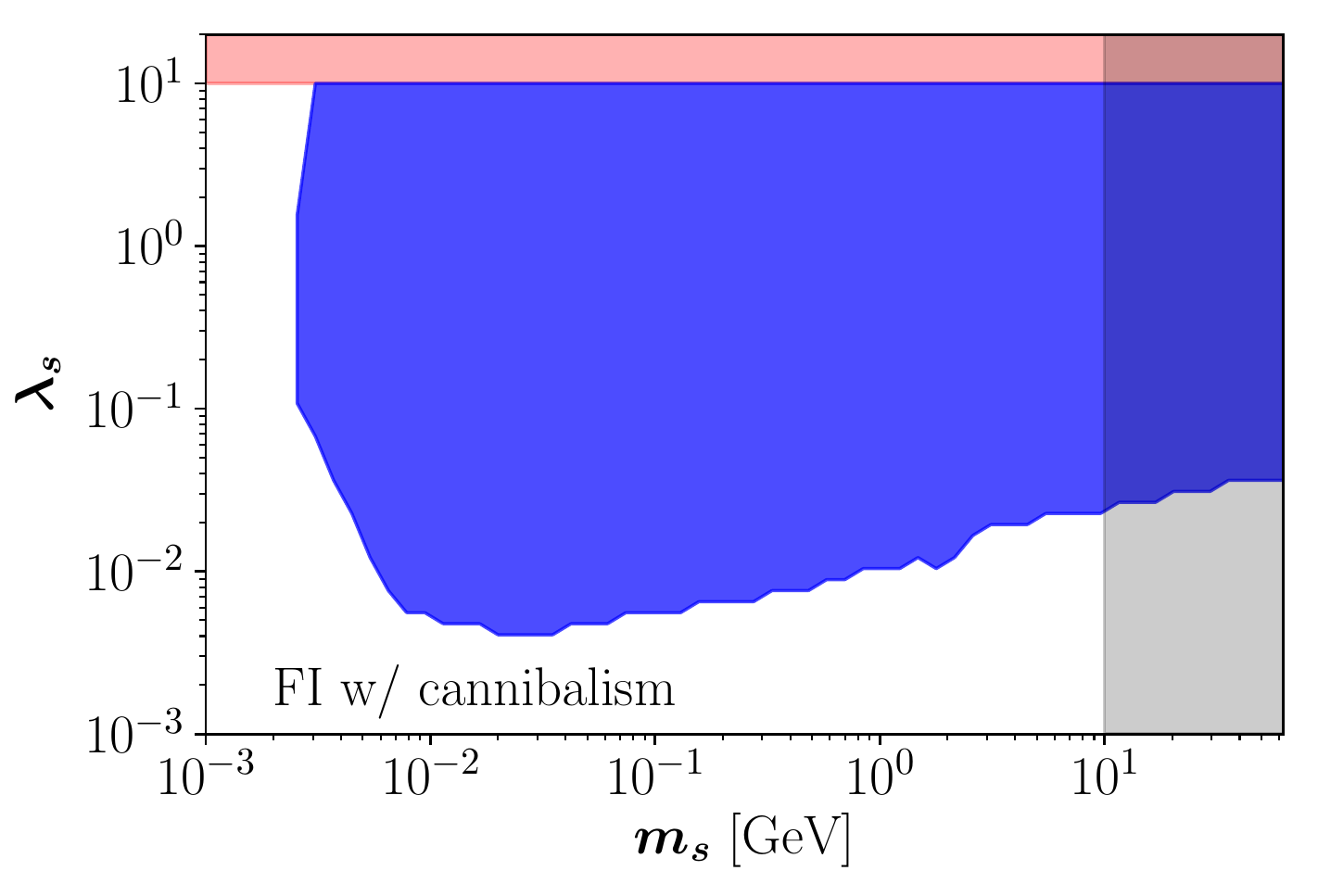}\\
\includegraphics[width=.46\textwidth]{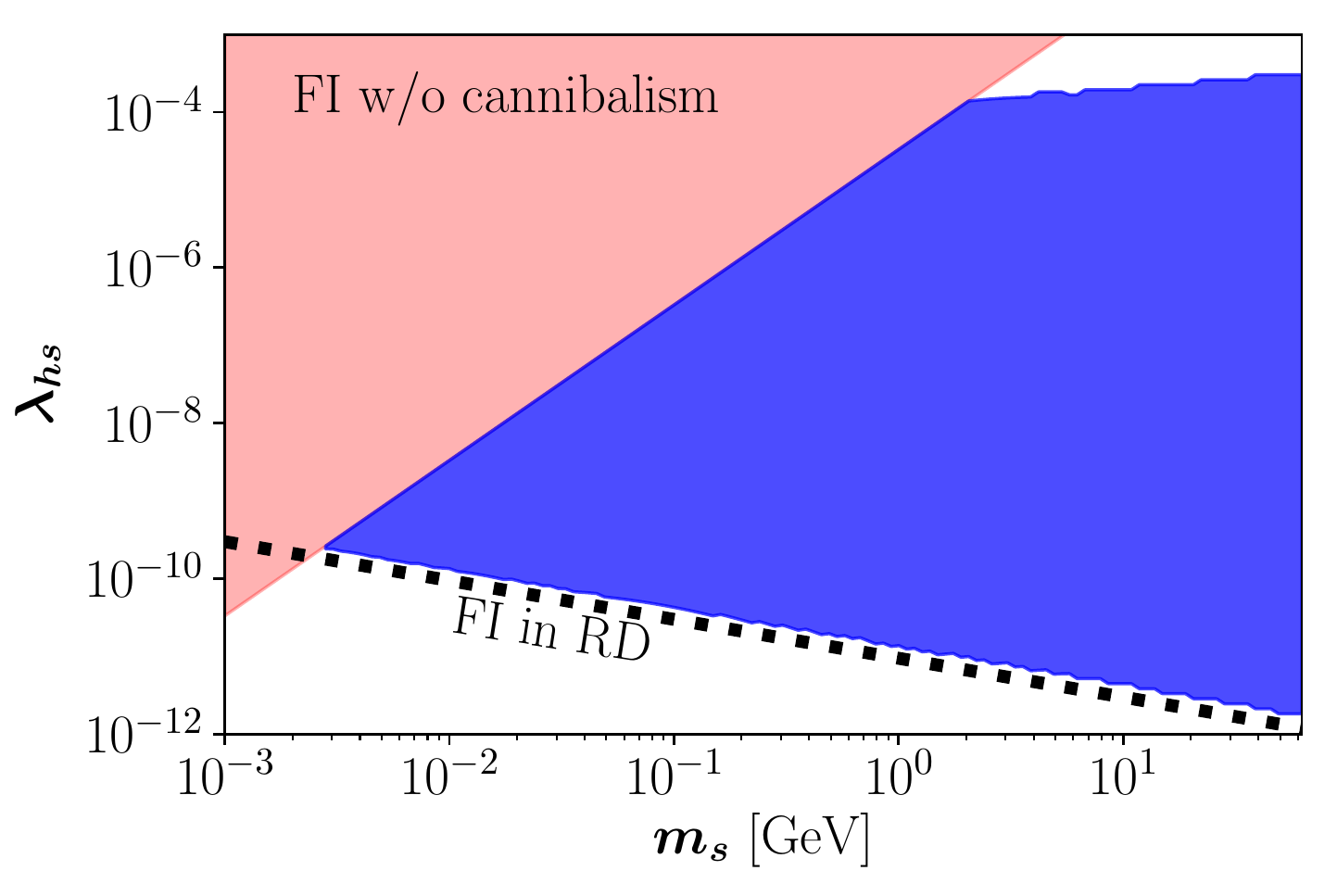}
\includegraphics[width=.46\textwidth]{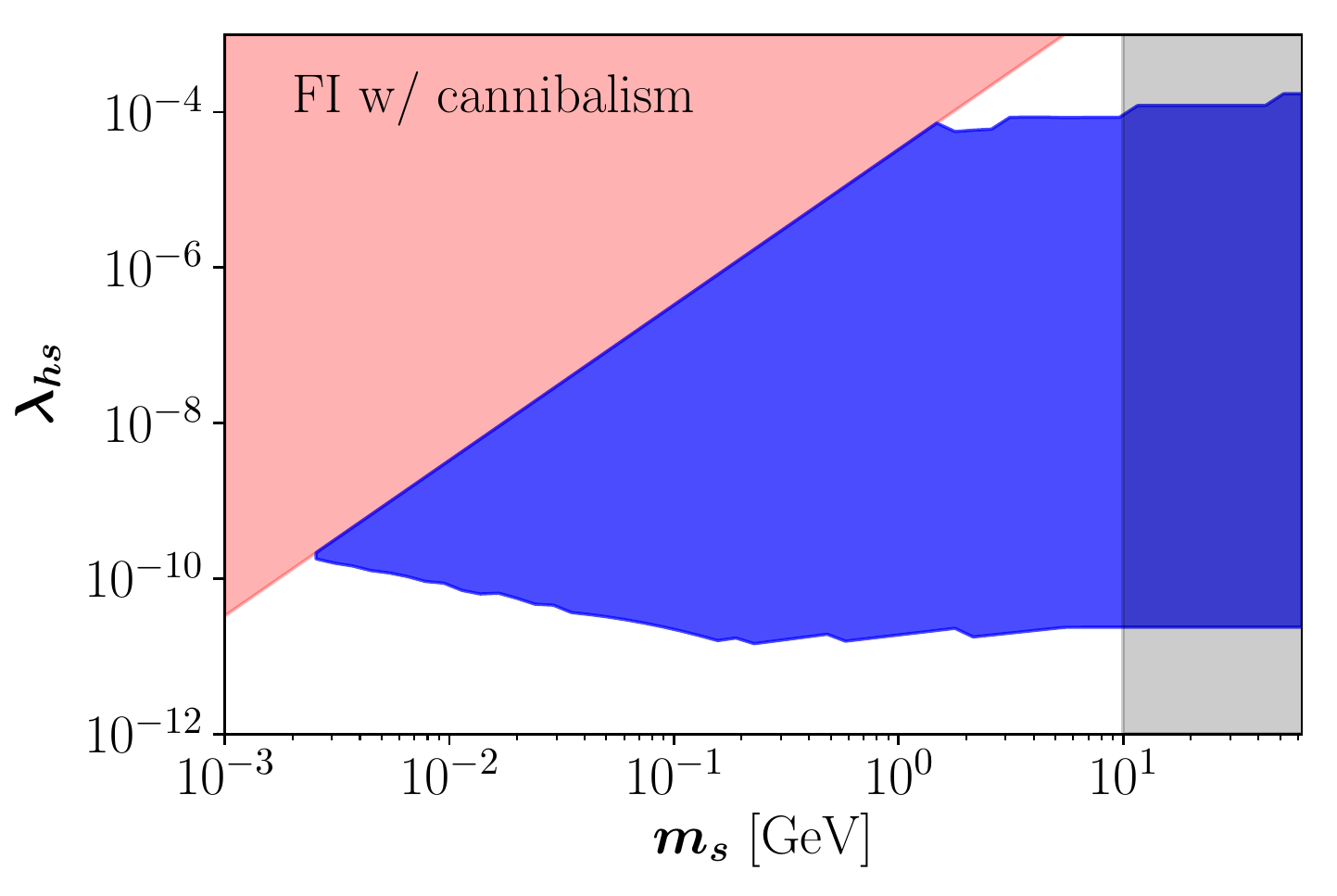}
\includegraphics[width=.46\textwidth]{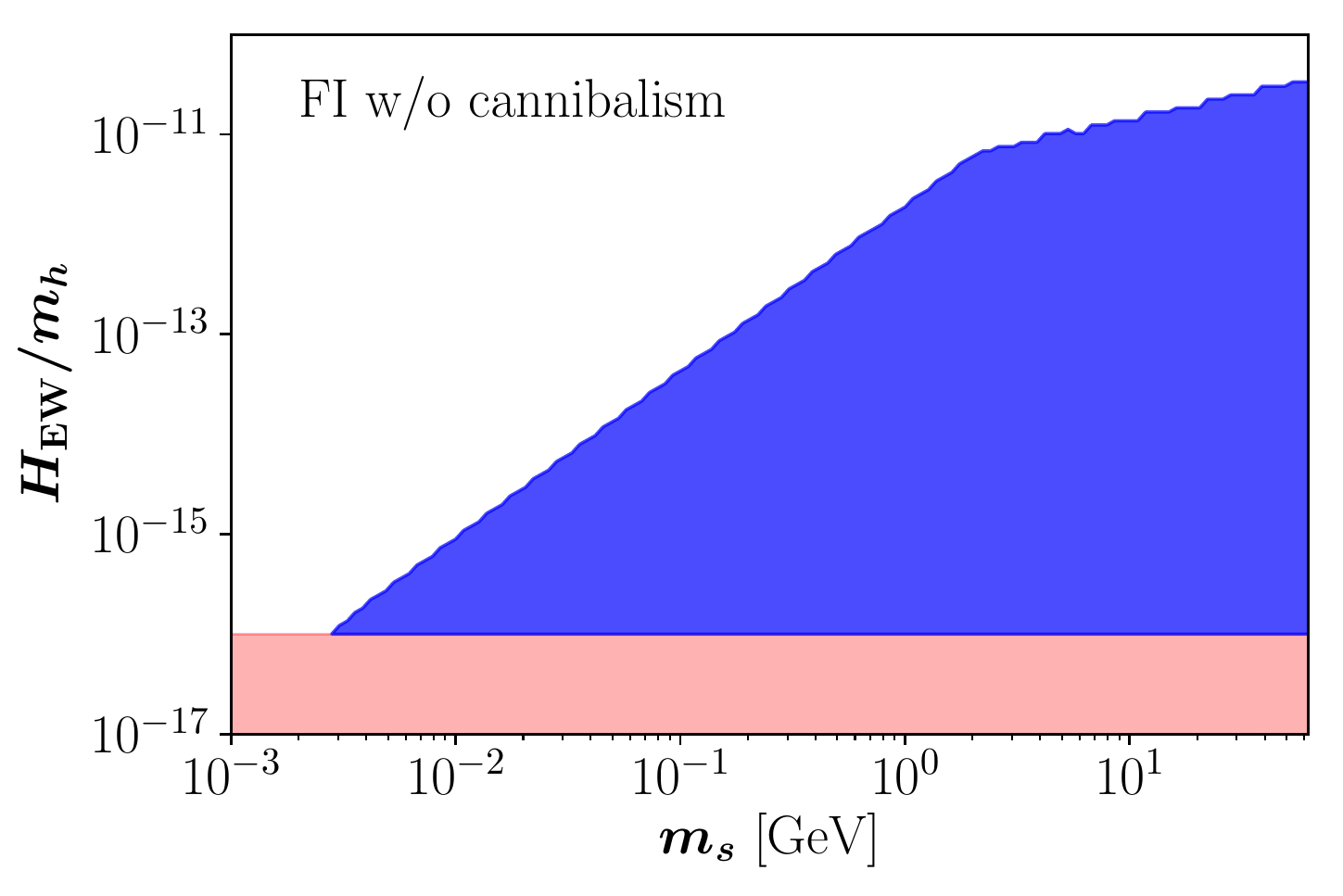}
\includegraphics[width=.46\textwidth]{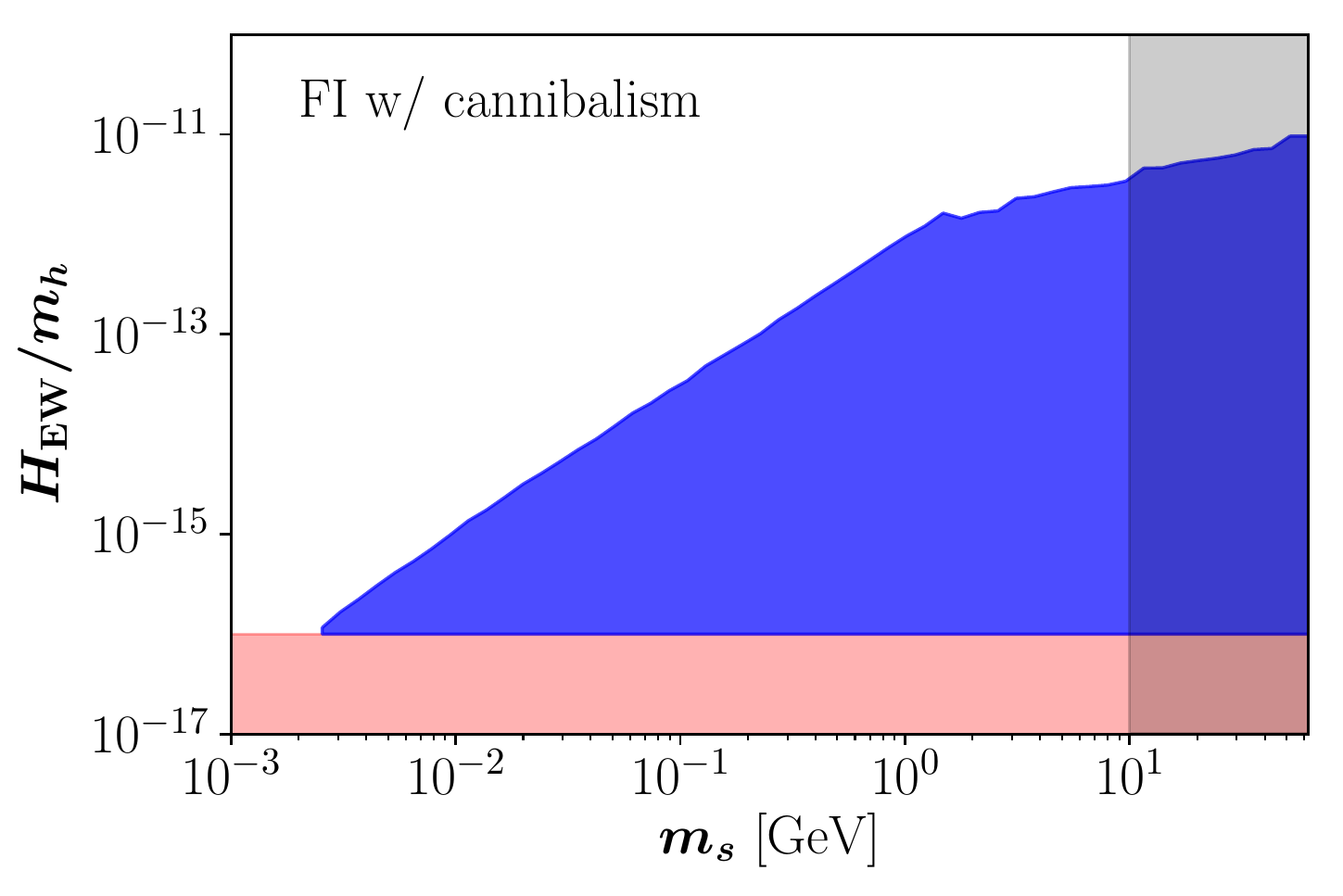}
\includegraphics[width=.46\textwidth]{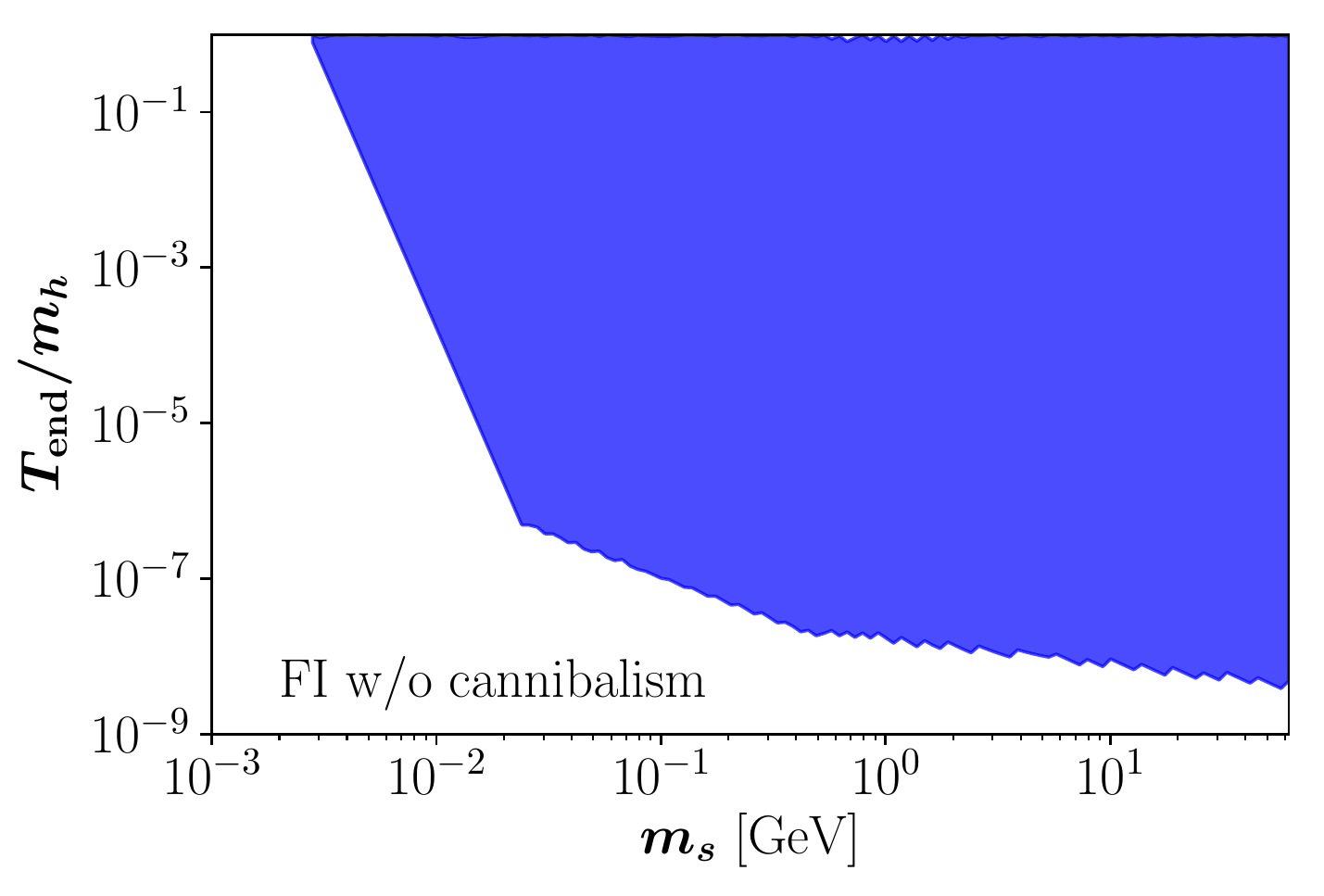}
\includegraphics[width=.46\textwidth]{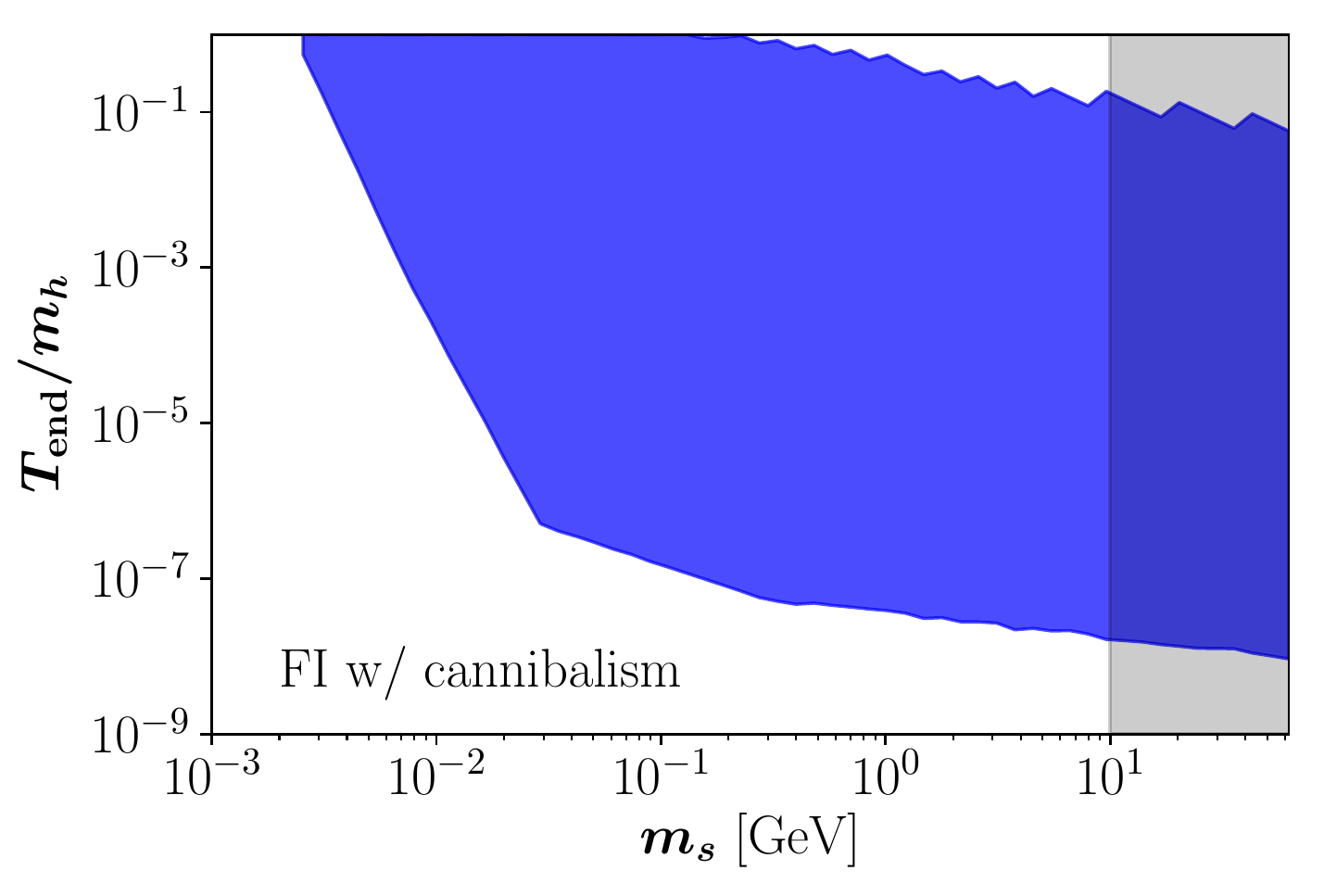}
\caption{DM freeze-in without (left column) and with (right column) cannibalism. Parameter space giving rise to the observed DM relic abundance. The black dotted line shows the parameters yielding the correct DM abundance in the usual RD scenario.
The red regions correspond to the constraints discussed in Section~\ref{sec:constraints}: the SM temperature after the matter-like component has decayed into SM particles must be larger than the BBN temperature and small enough not to not re-trigger DM production, Eq. (\ref{tend over mh constr}); in a MD Universe $H_{\mathrm{EW}}/m_h >1.76\times 10^{-16}$; the portal coupling has to satisfy $\lambda_{hs}<2\,m_s^2/v^2$ and $\lambda_{hs}<\lambda_{hs}^\text{eq}$ with $\lambda_{hs}^\text{eq}$ given by Eq.~\eqref{portalcoupling}. The shaded region in panels on the right hand side corresponds to the reannihilation regime. Other observational constraints are shown in Fig.~\ref{Detection}.}
\label{FI}
\end{center}
\end{figure*}
Fig.~\ref{FI} depicts the results of an extensive scan over the parameter space for the DM freeze-in without (left column) and with (right column) cannibalism. The blue regions produce the observed DM relic abundance, the red regions correspond to the constraints discussed in Section~\ref{sec:constraints}. Other observational constraints on the scenario will be discussed in Section~\ref{constraints}.

The plots generalize the results of Figs.~\ref{FIwithoutCannibalSlices} and~\ref{FIwithCannibalSlices}.
First, the usual RD scenario without cannibalism can be approximately recovered by taking $H_\text{EW}/m_h=H_\text{EW}^\text{rad}/m_h\simeq 1.76\times 10^{-16}$ and $T_\text{end}/m_h=1$, as discussed in Section \ref{sec:FI w/o cann}.
This corresponds to the black dotted line with $\lambda_{hs}\simeq\mathcal{O}(10^{-11})$.
Second, in the MD scenario the Higgs portal can reach much higher values up to $\mathcal{O}(10^{-4})$.
Such big values for freeze-in naturally need large dilution factors, characterized by large expansion rates $H_\text{EW}/m_h$ up to $\mathcal{O}(10^{-11})$ and low temperatures for the end of the MD era, $T_\text{end}/m_h$ down to $\mathcal{O}(10^{-8})$.
Higher values of $\lambda_{hs}$ cannot be reached, because in the present case thermalization with the SM must be avoided.


\section{Observational Properties}
\label{constraints}

Finally, we turn into observational prospects, discussing collider signatures, direct and indirect detection, as well as the observational consequences of DM self-interactions.

\subsection{Collider Signatures}

For small singlet masses, $m_s < m_h/2$, the Higgs can decay efficiently into a pair of DM particles. Thus, the current
limits on the invisible Higgs branching ratio (BR$_\text{inv}\lesssim 20\%$~\cite{Bechtle:2014ewa}) and the total Higgs decay width ($\Gamma^\text{tot}\lesssim 22$~MeV \cite{Khachatryan:2014iha}) constrain the Higgs portal coupling, $\lambda_{hs}$, by Eq. \eqref{Gamma}. This constraint applies to both freeze-out and freeze-in scenarios, although typically it can be expected to constrain only the freeze-out case, as usually in freeze-in scenarios the value of $\lambda_{hs}$ required to reproduce the observed DM abundance is orders of magnitudes below these values. Indeed, the collider signatures of frozen-in DM were recently deemed unobservable in Ref.~\cite{Kahlhoefer:2018xxo}. However, the paper considered only the usual RD case, and in a scenario containing an early phase of rapid expansion, such as in the present paper, the portal coupling can take a much larger value than what is usually encountered in the context of freeze-in. It is therefore not a priori clear whether constraints of the above kind can be neglected or not. We will present them in Section \ref{results}.

In MD cosmologies, the interaction rates required to produce the observed DM abundance via freeze-in could lead to displaced signals at the LHC and future colliders~\cite{Co:2015pka}.
However, as in our scenario DM is produced via the decay of the Higgs, we will have no exotic signals displaced from the primary vertex.

\subsection{Direct and Indirect Detection Signatures}

The direct detection constraint is obtained by comparing the spin-independent cross section for the scattering of the DM off of a nucleon,
\begin{equation}
\sigma_\text{SI}=\frac{\lambda_{hs}^2\,m_N^4\,f^2}{4\pi\,m_s^2\,m_h^4}\,,
\end{equation}
to the latest limits on $\sigma_\text{SI}$ provided by PandaX-II~\cite{Cui:2017nnn}, LUX~\cite{Akerib:2016vxi} and Xenon1T~\cite{Aprile:2017iyp}.
Here $m_N$ is the nucleon mass and $f\simeq 1/3$ corresponds to the form factor~\cite{Farina:2009ez,Giedt:2009mr,Alarcon:2011zs,Alarcon:2012nr}.
We also take into account the projected sensitivities of the next generation DM direct detection experiments like LZ~\cite{Akerib:2018lyp} and DARWIN~\cite{Aalbers:2016jon}.
Moreover, multiple experimental setups have recently been suggested for the detection of elastic scatterings of DM in the mass range from keV to MeV~\cite{Graham:2012su,Hochberg:2015pha,Essig:2015cda,Hochberg:2015fth,Hochberg:2016ajh,Schutz:2016tid,Derenzo:2016fse,Kouvaris:2016afs,Hochberg:2016sqx,Essig:2016crl,Knapen:2016cue,McCabe:2017rln,Budnik:2017sbu,Davis:2017noy,An:2017ojc}.
In particular, the typical DM-electron cross sections for MeV-scale FIMP DM could be tested by some next generation experiments~\cite{Essig:2011nj,Essig:2012yx,Essig:2017kqs,Dolan:2017xbu}.

The current limits from the analysis of gamma-rays coming from dwarf spheroidal galaxies with Fermi-LAT and DES~\cite{Fermi-LAT:2016uux,Benito:2016kyp,Calore:2018sdx} do not probe relevant parts of our parameter space. In the case of freeze-in, indirect detection signals can be expected in scenarios where the singlet scalar is a mediator and the hidden sector exhibits a richer structure, as recently studied in Ref.~\cite{Heikinheimo:2018duk}.

\subsection{Dark Matter Self-interactions}
Finally, we consider the observational ramifications of DM self-interactions. Two long-standing puzzles of the collisionless cold DM paradigm are the `cusp vs. core'~\cite{Moore:1994yx,Flores:1994gz,Navarro:1996gj,deBlok:2009sp,Oh:2010mc,Walker:2011zu} and the `too-big-to-fail'~\cite{BoylanKolchin:2011de,Garrison-Kimmel:2014vqa} problems.
These issues are collectively referred to as small scale structure problems of the $\Lambda$CDM model; for a recent review, see Ref.~\cite{Tulin:2017ara}.
These tensions can be alleviated if at the scale of dwarf galaxies DM exhibits a large self-scattering cross section, $\sigma$, over DM particle mass, $m_s$, in the range $0.1\lesssim\sigma/m_s\lesssim 10$~cm$^2$/g~\cite{Spergel:1999mh,Wandelt:2000ad,Buckley:2009in,Vogelsberger:2012ku,Rocha:2012jg,Peter:2012jh,Zavala:2012us,Vogelsberger:2014pda,Elbert:2014bma,Kaplinghat:2015aga}.
Nevertheless, the non observation of an offset between the mass distribution of DM and galaxies in the Bullet Cluster constrains such self-interacting cross section, concretely $\sigma/m_s < 1.25$~cm$^2$/g at $68\%$ CL~\cite{Clowe:2003tk,Markevitch:2003at,Randall:2007ph}. In the limit $m_s\ll m_h$ we have
\begin{equation}\label{sigma over m Bullet}
\frac{\sigma}{m_{s}}\simeq\frac{9}{32\pi}\,\frac{\lambda_{s}^{2}}{m_{s}^{3}}\lesssim1.25\,\mathrm{\frac{cm^{2}}{g}}\,, 
\end{equation}
which imposes an important constraint
\begin{equation}
\lambda_{s}\lesssim2\times10^{2}\,\left(\frac{m_{s}}{\mathrm{GeV}}\right)^{3/2}\,,
\label{lambda s constr}
\end{equation}
which we will show in our results in the next Subsection.

In the present case, no cosmological signatures can be expected. Even though in the case where the singlet scalar never thermalizes with the SM sector the DM generically comprises an isocurvature mode in the CMB fluctuations \cite{Nurmi:2015ema,Kainulainen:2016vzv}, the relative amount of such perturbations gets strongly diluted due to $\rho_s \ll \rho_{\rm tot}$, leaving no observable imprints on the CMB.

\subsection{Results}
\label{results}

\begin{figure*}
\begin{center}
\hspace{7.5cm}
\includegraphics[width=.42\textwidth]{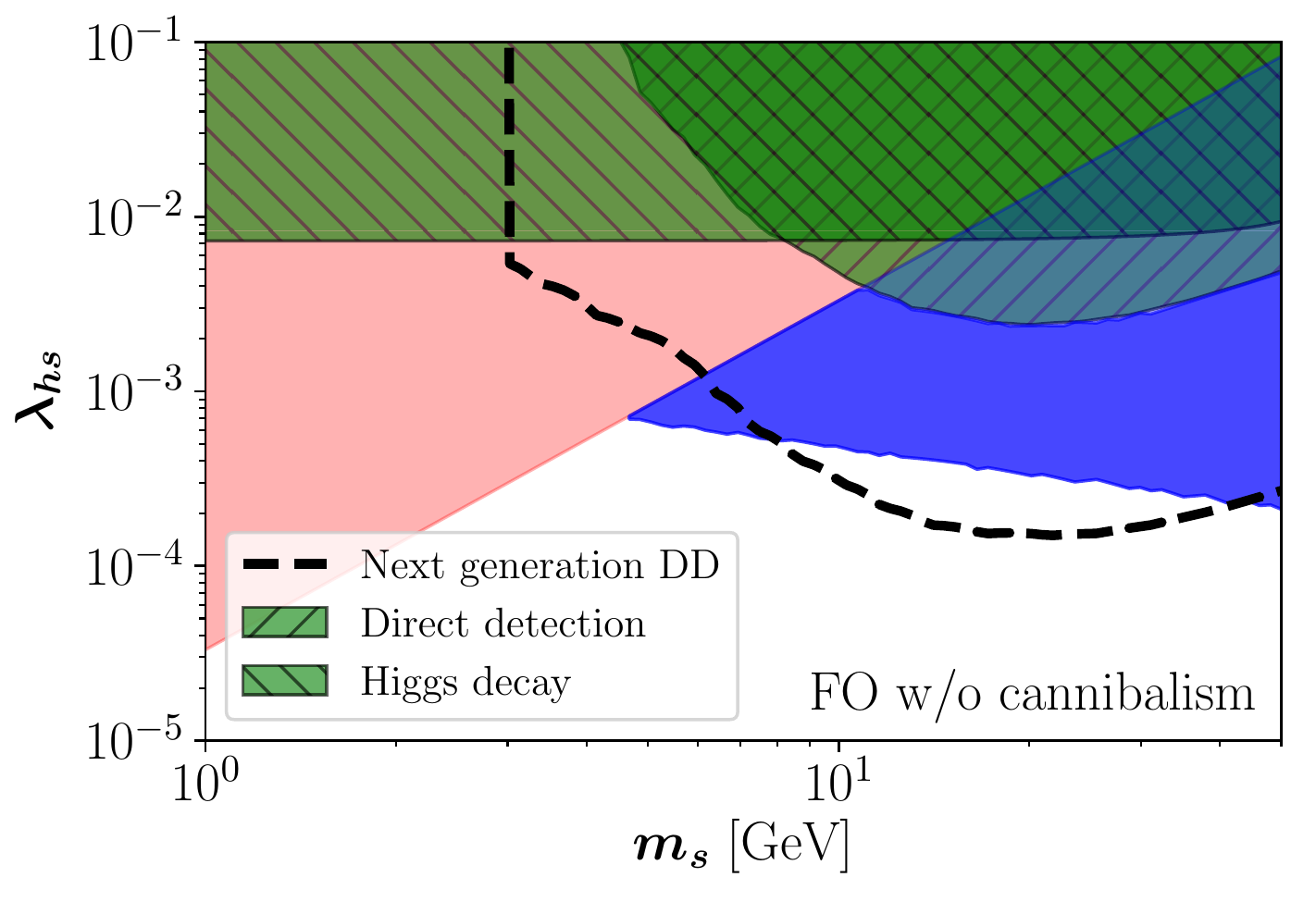}\\
\includegraphics[width=.42\textwidth]{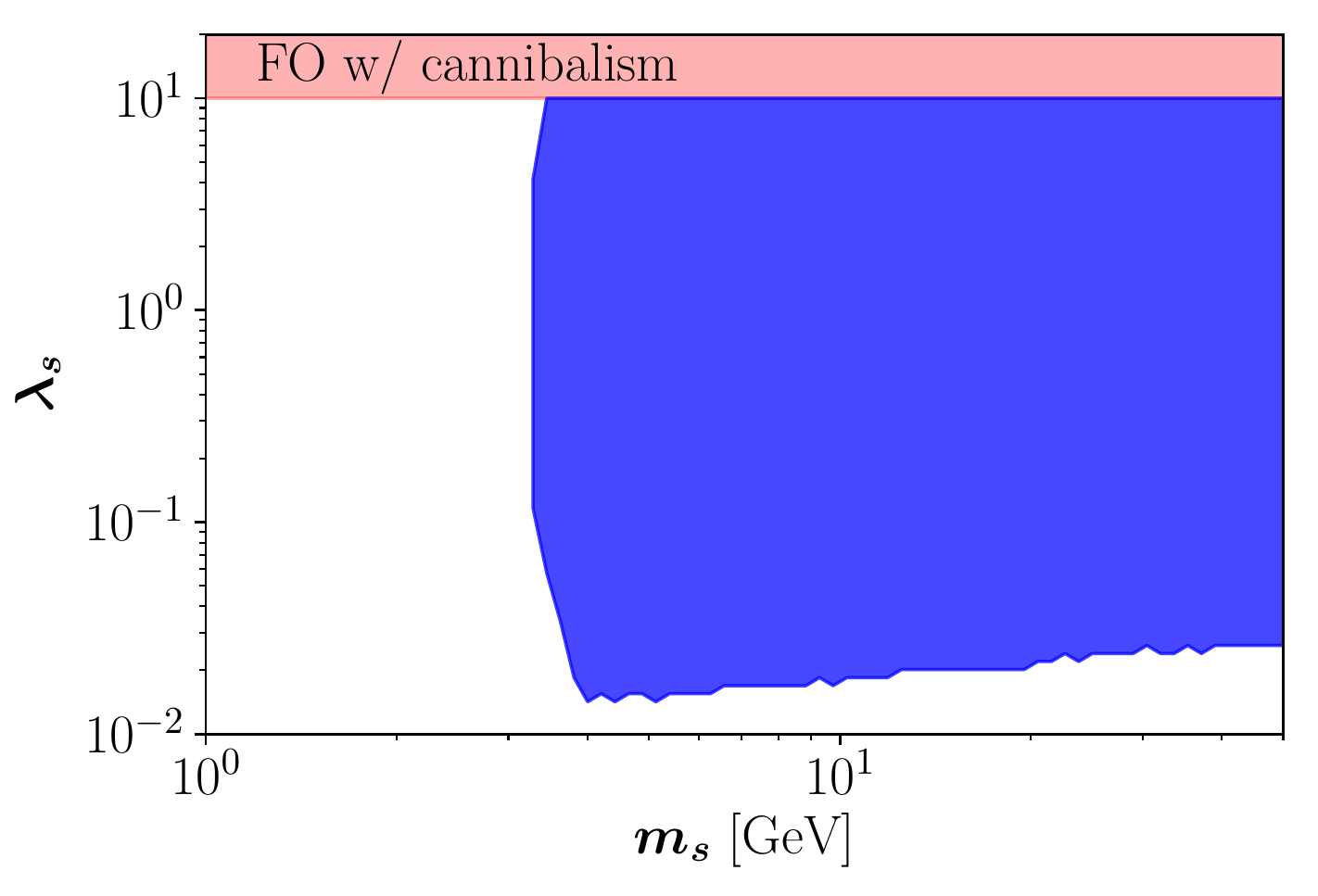}
\includegraphics[width=.42\textwidth]{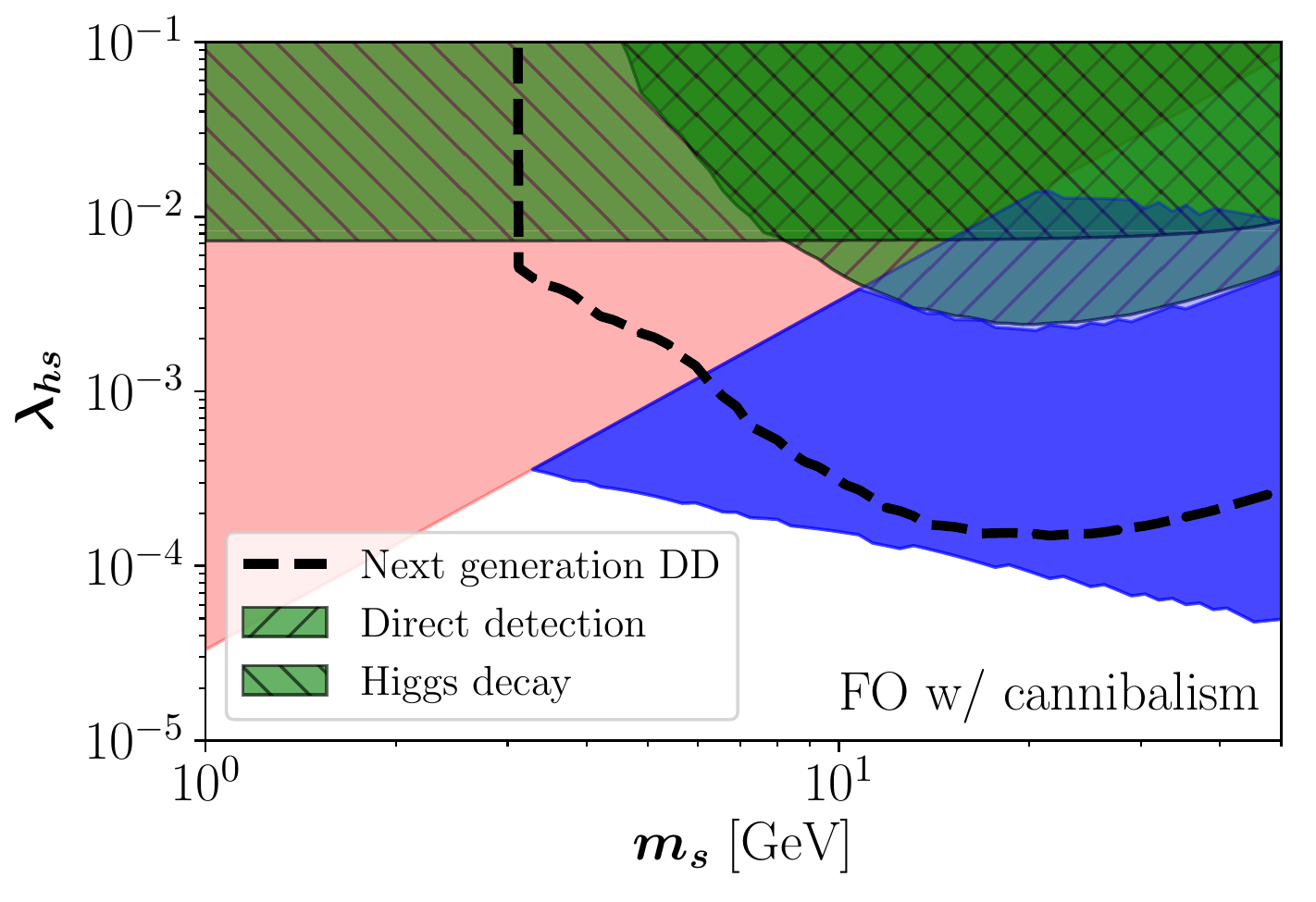}\\
\hspace{7.5cm}
\includegraphics[width=.42\textwidth]{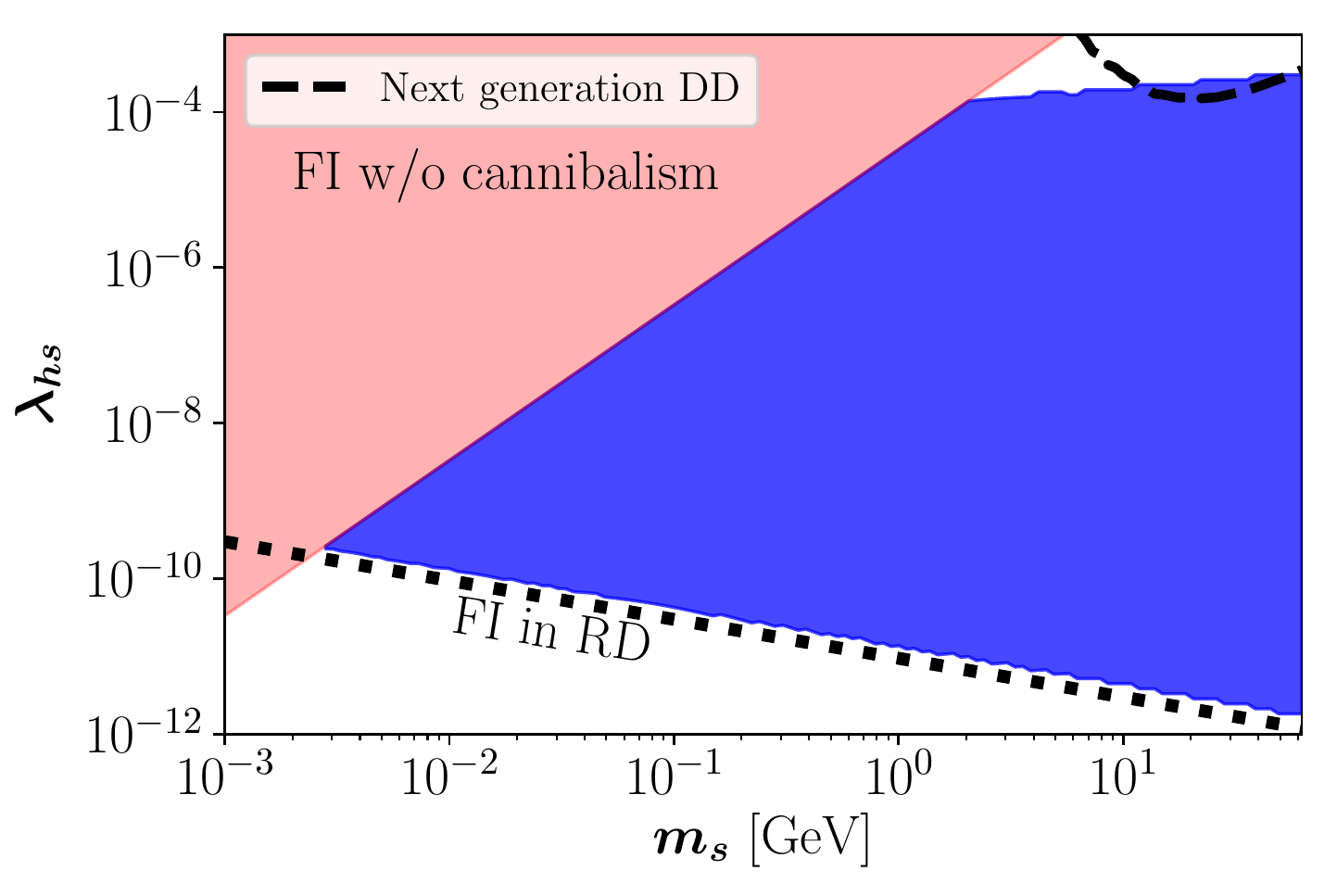}\\
\includegraphics[width=.42\textwidth]{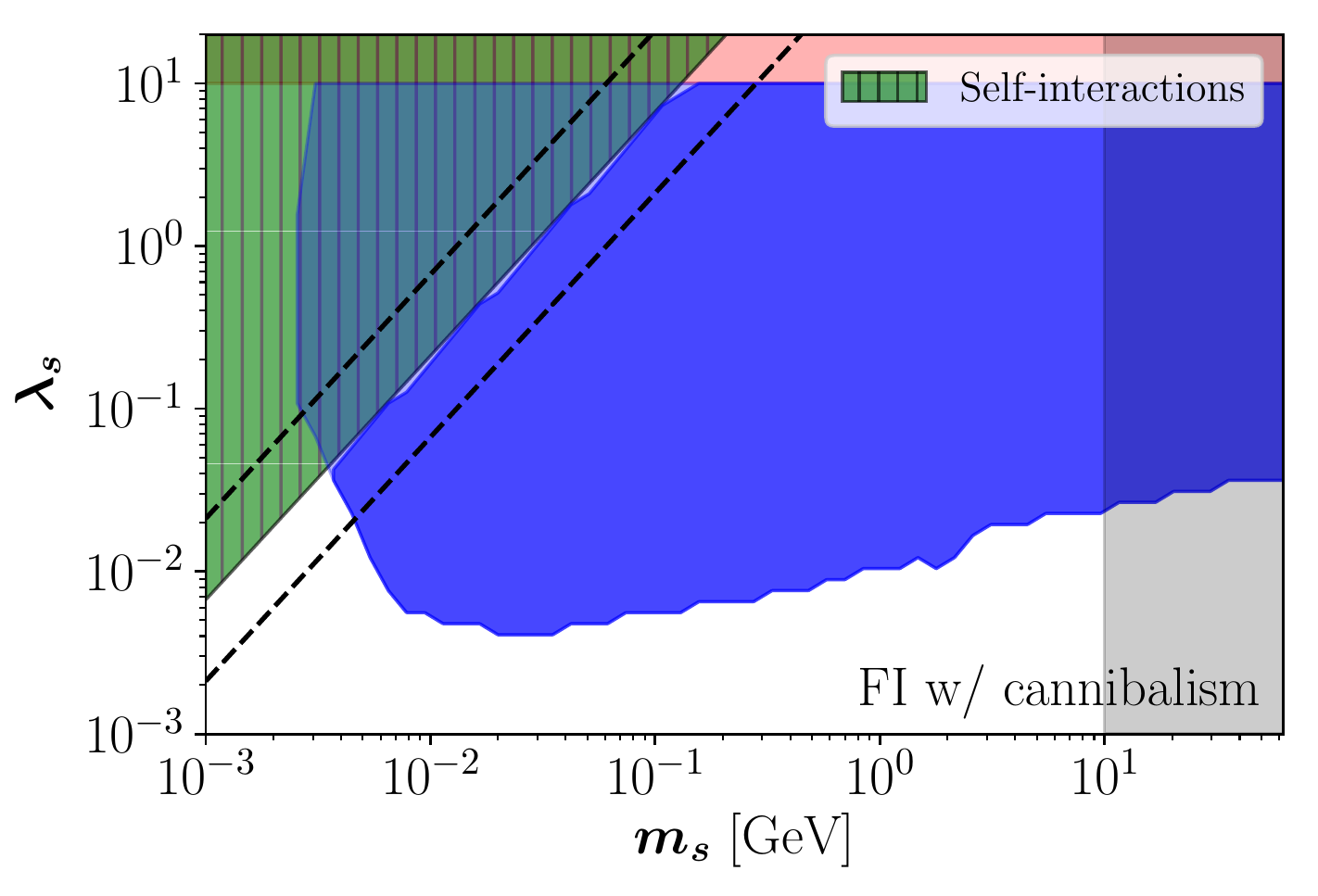}
\includegraphics[width=.42\textwidth]{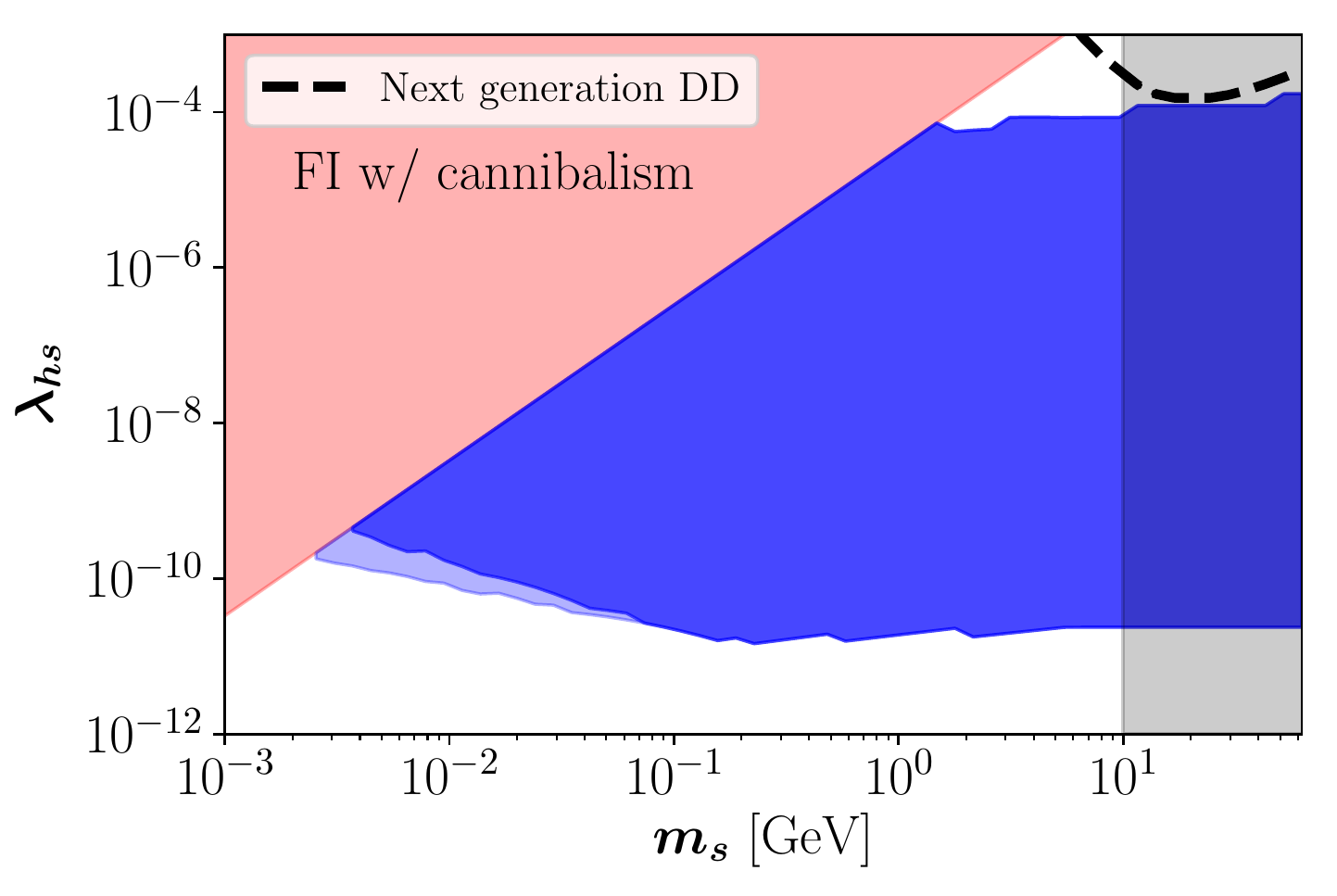}
\caption{Detection prospects for frozen-out and frozen-in DM with and without cannibalism, as indicated in the figures.
The green regions are excluded by different measurements: DM direct detection, invisible Higgs decay, or DM self-interactions.
The blue regions give rise to the observed DM relic abundance, the light blue being already in tension with observations.
The black thick dashed line corresponds to the bounds that might be reached by next generation direct detection DM experiments.
The red regions correspond to the constraints discussed in Section~\ref{sec:constraints}: the SM temperature after the matter-like component has decayed into SM particles must be larger than the BBN temperature and small enough not to not re-trigger DM production, Eq. (\ref{tend over mh constr}); the DM freeze-out occurs while the $s$ particles are non-relativistic, $x_{\rm FO}>3$; in a MD Universe $H_{\mathrm{EW}}/m_h >1.76\times 10^{-16}$; the portal coupling has to satisfy $\lambda_{hs}<2\,m_s^2/v^2$ and $\lambda_{hs}\geq \lambda_{hs}^\text{eq}$ for the freeze-out case and $\lambda_{hs}< \lambda_{hs}^\text{eq}$ for the freeze-in case, with $\lambda_{hs}^\text{eq}$ given by Eq.~\eqref{portalcoupling}. The shaded region in the lower panels corresponds to the reannihilation regime. The black dotted line shows the parameters yielding the correct DM abundance in the usual RD scenario.}
\label{Detection}
\end{center}
\end{figure*}
Fig.~\ref{Detection} depicts the detection prospects for frozen-out and frozen-in DM, with and without cannibalism.
The green regions are excluded by different observations discussed in the above subsections: DM direct detection, invisible Higgs decay or DM self-interactions.
The blue regions give rise to the observed DM relic abundance, the light blue region being already in tension with observations.
The black thick dashed line corresponds to the bounds that might be reached by next generation direct detection DM experiments.
The constraints discussed in Section~\ref{sec:constraints} are shown in red.
Finally, the black dotted line shows the parameters yielding the correct DM abundance in the usual RD scenario.

In the MD scenario, DM direct detection already excludes an important region of the parameter space for the freeze-out case both with and without cannibalism. More interestingly, the next generation of DM direct detection experiments will be able to probe almost the whole region of parameter space compatible with the DM relic abundance, for the freeze-out scenario with $m_s<m_h/2$.

On the other hand, the regions favored by freeze-in could be tangentially probed by next generation of direct detection experiments. A particularly interesting thing in this case is that the effect of non-vanishing self-interactions seems to be crucial in determining whether the scenario can be tested by the next-generation direct detection experiments or not, as shown in the two lower panels of Fig.~\ref{Detection}. In the standard RD case the freeze-in scenario obviously does not have any such observational consequences, as the required values of $\lambda_{hs}$ are in that case much smaller regardless of the value of $\lambda_s$. Note that in Ref.~\cite{Bernal:2018kcw} we obtained the opposite result, showing that FIMP DM cannot be tested by the next-generation direct detection experiments. This conclusion, however, is due to different assumptions for the decay of the matter-like component, as discussed in Section~\ref{intro}. However, observational constraints on DM self-interactions already rule out a corner of the parameter space corresponding to MeV-scale masses regardless of the prospects for direct detection. Finally, the region between the two dashed lines in the case of freeze-in with cannibalism corresponds to $0.1$~cm$^2$/g$~<\sigma/m_s<10$~cm$^2$/g, the zone where the small-scale structure tensions can be alleviated.


\section{Conclusions}
\label{conclusions}

In cosmology, one typically assumes that at early times the Universe was radiation-dominated from the end of inflation. However, there are no indispensable reasons to assume that, and alternative cosmologies not only can lead to interesting observational ramifications but are also well-motivated. For example, an early period of matter domination is still a perfectly viable option.

In this context, we studied different dark matter production mechanisms during an early MD era.
We focused first on the usual case where DM is produced by the freeze-out mechanism, corresponding to the WIMP paradigm.
Then, the assumption of thermal equilibrium with the SM was relaxed allowing the DM to be produced via the freeze-in mechanism, corresponding to FIMP DM. For these two cases, we took for the first time into account the effects of sizable self-interactions within the hidden sector. Indeed, as we showed in the present context, DM self-interactions can be crucial for the determination of the final DM relic abundance and observational consequences.

When the expansion rate of the Universe differs from the usual radiation-dominated case, it tends
to effectively dilute the DM abundance when the era of non-standard expansion ends and the
visible sector gets reheated. This means that in case the expansion was faster than in the RD case and the DM particles were initially in thermal equilibrium with the visible sector, they generically have to undergo freeze-out earlier than
in the usual RD case, thus resulting in larger DM abundance to match the observed one. In
case the DM particles interacted so feebly that they never became part of the SM equilibrium heat
bath, the coupling between DM and the visible sector typically has to be orders of magnitude
larger than in the usual freeze-in case to compensate the larger expansion rate. As we showed, sizable self-interactions can further complicate this picture. Production of self-interacting DM during a non-standard expansion phase may thus result in important experimental and observational ramifications, as shown in Fig. \ref{Detection}.

In this paper we studied a benchmark scenario where the SM is extended with a real singlet scalar DM, odd under a $\mathbb{Z}_2$ symmetry. It would be interesting to see what are the consequences in other models where, for example, the hidden sector has a richer structure (e.g. sterile neutrinos, gauge structure, etc.) or where the DM is not coupled to the SM via the Higgs portal but via some other portal, for example the $Z'$  or a lepton portal~\cite{Pospelov:2007mp,Krolikowski:2008qa,Bai:2014osa}.


\section*{Acknowledgments}
We thank X. Chu, M. Heikinheimo, M. Lewicki, and V. Vaskonen for correspondence and discussions. T.T. acknowledges Universidad Antonio Nari\~{n}o and Universidade do Porto for hospitality. C.C. is supported by the Funda\c{c}\~{a}o para a Ci\^{e}ncia e Tecnologia (FCT) grant\linebreak PD/BD/114453/2016, T.T. by the Simons foundation and the U.K. Science and Technology Facilities Council grant ST/J001546/1, and N.B. partially by Spanish MINECO under Grant FPA2017-84543-P. This project has also received funding from the European Union’s Horizon 2020 research and innovation programme under the Marie\linebreak Skłodowska-Curie grant agreements 674896 and 690575; and from Universidad Antonio Nariño grants 2017239 and 2018204.


\appendix
\section{Dark Matter Abundance in the Present Universe}
\label{appendix}

The DM abundance at present is
\begin{equation}
\label{present abundance}
\Omega_{s}h^2= \frac{\rho_s}{\rho_{\rm c}/h^2} = \frac{\xi\, \mathfrak{s}_0}{\rho_{\rm c}/h^2} \,,
\end{equation}
where $\mathfrak{s}_0=2891$~cm$^{-3}$ and $\rho_{\rm c}/h^2 = 1.054\times 10^{-5}$~GeV/cm$^3$ are, respectively, the entropy density and critical energy density today~\cite{Co:2015pka}, and
\be
\label{xi}
\xi \equiv \frac{\rho_s(T_{\rm end}')}{\mathfrak{s}(T_{\rm end}')} = m_s\frac{n_s(T_{\rm end}')}{\mathfrak{s}(T_{\rm end}')} = m_s\frac{\chi_s^\infty}{\mathcal{S}(T_{\rm end}')},
\ee
where $\chi_s^\infty\equiv a^3 n_s$ is the comoving DM number density after freeze-in/-out and the SM entropy at the temperature the SM sector gained when the MD ended, $T_{\rm end}'$, is given by
\be
\label{S(T')}
\mathcal{S}(T_{\rm end}') = \frac{2\pi^2}{45}g_{*\mathfrak{s}}(T_{\rm end}')\,T_{\rm end}'^3\, a_{\rm end}^3 \,.
\ee
Only after this point the comoving entropy density in the SM sector is conserved. Note that from this point on, the expansion history of the Universe does not affect the result. In Eq.~\eqref{S(T')},  $a_{\rm end}$ can be replaced by $H_{\rm EW}$ by using the Friedmann equation, $H_{\rm end} \propto H_{\rm EW}\,a_{\rm end}^{-3/2} \propto T_{\rm end}'^2/M_{\rm P}$, so that
\be
a_{\rm end}^3 = \left(\frac{90}{\pi^2 g_*(T_{\rm end}')}\right)\left(\frac{M_{\rm P}\,H_{\rm EW}}{T_{\rm end}'^2}\right)^2 \,.
\ee
We reiterate that we have normalized the scale factor so that $a_{\rm EW}=1$.

One can then either substitute the comoving number density $\chi_s^\infty$ into Eq.~\eqref{xi} (as in the case of Eq. \eqref{chi_infty}, which gives the result \eqref{abundance_notherm}) or calculate the actual DM number density $n_s(T_{\rm end}')$ in Eq. \eqref{xi} by relating it to the number density at the time the DM production ended
\begin{equation}
n_{s}\left(T_{\rm end}'\right)=n_{s}^{\mathrm{final}}\left(T_{\mathrm{F}}\right)\frac{g_{*\mathfrak{s}}(T_{\rm end})}{g_{*\mathfrak{s}}(T_{\rm F})}\,\left(\frac{T_{\mathrm{end}}}{T_{\rm F}}\right)^{3},\label{ns at Tend}
\end{equation}
as in the case of Eqs.~\eqref{FOresult}, \eqref{nsFO} and~\eqref{ns final}.
Relating $n_{s}\left(T_{\rm end}'\right)$ to $T_{\rm end}$ but using $T_{\rm end}'$ for the entropy density $\mathfrak{s}$ in Eq. \eqref{xi} leads to an artificial discontinuity in DM number density. This reflects the fact that we assume that the dominant matter-like component decays instantaneously to the SM sector, heating the SM particles instantaneously from temperature $T_{\rm end}$ to a higher temperature $T_{\rm end}'$ and simultaneously effectively diluting the DM number density.

The relation between $T_{\mathrm{end}}$ and $T_{\mathrm{end}}'$ can be found as follows. Following Eq. \eqref{Hubble}, the matter-like component's energy density can be written as
\begin{equation}
\rho_{\mathrm{M}}\left(T\right)=3\,M_{\mathrm{P}}^{2}\,H_{\mathrm{EW}}^{2}\,\left(\frac{T}{m_h}\right)^{3}\,\left(\frac{g_{*}\left(T\right)}{g_{*}\left(m_h\right)}\right),\label{matter density}
\end{equation}
and the SM energy density in the usual way as
\begin{equation}
\rho_{\mathrm{SM}}\left(T\right)=\frac{\pi^{2}}{30}\,g_{*}\left(T\right)\,T^{4}.\label{radiation density}
\end{equation}
At $T=T_{\mathrm{end}}$, the matter-like component transfers all of its energy into the SM sector,
$\rho_{\mathrm{M}}\left(T_{\mathrm{end}}\right)=\rho_{\mathrm{SM}}\left(T_{\mathrm{end}}'\right)$, so that one finds
\begin{align}
	\frac{T_{\mathrm{end}}}{T_{\mathrm{end}}'} \simeq 0.4\left(\frac{H_{\rm EW}/m_h}{10^{-16}}\right)^{-1/2}\,\left(\frac{T_{\mathrm{end}}}{m_h}\right)^{1/4}\nonumber\\
	\times \left(\frac{g_{*}\left(T_{\mathrm{end}}'\right)}{g_{*}\left(T_{\mathrm{end}}\right)}\,g_{*}\left(m_h\right)\right)^{1/4}.\label{Tend over Tend prime}
\end{align}
Substituting this results into Eq. \eqref{ns at Tend} and the resulting expression into Eq. \eqref{xi} then gives the present DM abundance as a function of the model parameters. This procedure gives us the results~\eqref{nsFO} and~\eqref{ns final}.

The relation \eqref{Tend over Tend prime} also makes it possible to constraint the duration of the early MD phase. As discussed in Section \ref{sec:constraints}, we require that the SM temperature after the matter-like component has decayed into SM particles, $T_{\rm end}'$, must be larger than the BBN temperature $T_{\rm BBN}=4$ MeV, and also that the temperature has to be smaller than either the final freeze-out temperature or smaller than $m_h$ in the freeze-in case in order not to re-trigger the DM yield after the decay of the matter-like component. This is what gives the conditions in Eq.~\eqref{tend over mh constr}. In order to determine the numerical values, we use $g_*(T_{\rm end}')=106.75$ for the upper limit and $g_*(T_{\rm end}')=10.75$ for the lower limit.


\bibliography{FIMPpheno}


\end{document}